\documentclass[12pt, twocolumn,showpacs,preprintnumbers,aps,prl,reprint,superscriptaddress]{revtex4-1}

\usepackage{graphicx}
\usepackage{subfigure}
\usepackage{color}
\usepackage{amsmath}
\usepackage[linktocpage,colorlinks=true,linkcolor=blue,citecolor=blue,breaklinks=true]{hyperref}
\usepackage{breakcites}
\usepackage{pgfplots}
\usepackage{natbib}
\usepackage{nameref}
\usepackage{xcolor}
\usepackage{cleveref}
\newcommand{\be}{\begin{equation}}
\newcommand{\ee}{\end{equation}}

\begin{document}

\preprint{}

\title{Thermal Convection over Fractal Surfaces}

\author{Srikanth Toppaladoddi}
\affiliation{University of Oxford, Oxford OX1 3PU, U.K.}

\author{Andrew J. Wells}
\affiliation{University of Oxford, Oxford OX1 3PU, U.K.}

\author{Charles R. Doering}
\affiliation{University of Michigan, Ann Arbor, MI 48109-1042, U.S.A.}

\author{John S. Wettlaufer}
\affiliation{Yale University, New Haven, CT 06520-8109, U.S.A.}
\affiliation{Nordita, Royal Institute of Technology and Stockholm University, Stockholm 106 91, Sweden}
\email[]{john.wettlaufer@yale.edu}

\date{\today}

\begin{abstract}
{We use well resolved numerical simulations with the Lattice Boltzmann Method to study Rayleigh-B\'enard convection in cells with a fractal boundary in two dimensions for $Pr = 1$ and $Ra \in \left[10^7, 10^{10}\right]$. The fractal boundaries are functions characterized by power spectral densities $S(k)$ that decay with wavenumber, $k$, as $S(k) \sim k^{p}$ ($p < 0$). The degree of roughness is quantified by the exponent $p$ with $p < -3$ for smooth (differentiable) surfaces and $-3 \le p < -1$ for rough surfaces with Hausdorff dimension $D_f=\frac{1}{2}(p+5)$. By computing the exponent $\beta$ in power law fits $Nu \sim Ra^{\beta}$, where $Nu$ and $Ra$ are the Nusselt and the Rayleigh numbers for $Ra \in \left[10^8, 10^{10}\right]$, we observe that heat transport scaling increases with roughness over the top two decades of $Ra \in \left[10^8, 10^{10}\right]$. For $p$ $= -3.0$,  $-2.0$ and $-1.5$ we find $\beta = 0.288 \pm 0.005, 0.329 \pm 0.006$ and $0.352 \pm 0.011$, respectively. We also observe that the Reynolds number, $Re$, scales as $Re \sim Ra^{\xi}$, where $\xi \approx 0.57$ over $Ra \in \left[10^7, 10^{10}\right]$, for all $p$ used in the study. For a given value of $p$, the averaged $Nu$ and $Re$ are insensitive to the specific realization of the roughness.}
\end{abstract}


\maketitle

\section{Introduction} 
Thermal convection refers to fluid flows that are driven by buoyancy forces due to density variations, which in turn are effected by gradients in temperature \citep{chandra2013}.
Such flows are ubiquitous in both the natural and engineering environments, and are key to understanding transport phenomena in the atmospheric boundary layer, in the outer core of Earth, and in the outer layers of stars \citep{kadanoff2001, wettlaufer2011} to name a few examples.
The simplest setting in which thermal convection can be studied is classical Rayleigh-B\'enard convection (RBC) in which a fluid is confined between two flat horizontal plates with the under side maintained at a higher temperature than the top \citep{Rayleigh1916}. Applying the Boussinesq approximation to the Navier-Stokes equations, the dynamics of RBC are governed by three non-dimensional parameters: the Rayleigh number $Ra$, the ratio of buoyancy to viscous forces, the Prandtl number $Pr$, the ratio of the fluid's kinematic viscosity to its thermal diffusivity, and the aspect ratio $\Gamma$ of the flow domain.

Heat transport in a fluid at rest is due solely to thermal conduction and when convective motions ensue this transport is enhanced.
The Nusselt number $Nu$, the ratio of total heat flux to conductive heat flux, is the quantitative measure of this enhancement.
Determining the dependence of $Nu$ on $Ra$, $Pr$, and $\Gamma$ for asymptotically large values of $Ra$ has been a major goal of the studies of convection; see, e.g., \citep{spiegel1971, kadanoff2001, ahlers2009, schumacher2012} and references therein.
Specifically, if $Nu$ is sought in terms of a power-law $Nu = A(Pr, \Gamma) \, Ra^{\beta}$ then the goal is to determine the value of the exponent $\beta$ for $Ra \gg 1$.

For planar geometries, if one assumes that the dimensional heat flux becomes independent of the depth of the cell as $Ra \rightarrow \infty$, then one obtains $Nu \sim Ra^{1/3}$.
This is the so-called {\it classical theory} of \citet{priestley1954}, \citet{malkus1954} and \citet{howard1966}.
However, if one assumes that the dimensional heat flux becomes independent of the molecular properties of the fluid when $Ra \rightarrow \infty$, then one obtains $Nu \sim \left(Pr \, Ra\right)^{1/2}$.
This `mixing length'  theory is originally due to \citet{Spiegel1963} and such scaling behavior---with possible logarithmic corrections \citep{kraichnan1962,chavanne1997}---is now often referred to as the {\it ultimate regime} of thermal convection.
The scaling $Nu \sim Ra^{1/2}$ is also an upper limit (uniformly in $Pr$) to the asymptotic heat transport scaling as $Ra \rightarrow \infty$ for no-slip fixed-temperature boundaries whether they are flat \citep{Howard63, Doering:1996} or corrugated, i.e., textured but sufficiently smooth  \citep{GD2016}.
(For flat no-slip boundaries at infinite Prandtl number the best known upper bound corresponds to the classical scaling $Nu \lesssim Ra^{1/3}$ within logarithmic corrections \citep{CD1999,DOR2006,OS2011}.)
In a wide range of studies at ${\cal O}(1)$ Prandtl number, the exponent $\beta$ is found to vary between $2/7$ \citep{urban2011, urban2012, verzicco2003, doering2009, stevens2010, DTW2019, iyer2020} and $1/3$ \citep{sreenivasan2000, verzicco2003, niemela2003, niemela2006, stevens2010, urban2011, urban2012, DTW2019, iyer2020}.
Several experiments have reported $\beta > 1/3$ \citep{chavanne1997, he2012}; however, because of the diversity of scalings reported for overlapping ranges of $Ra$, those findings await independent confirmation \citep{urban2012, he2013, urban2013, skrbek2015, he2016}.

The key difference between the classical ($\beta = 1/3$) and the ultimate ($\beta = 1/2$) theories principally lies in the role played by the thermal boundary layers.
In the former regime, thermal boundary layers presumably limit the rate of transport and hence control it \citep{howard1966}.
In the latter regime, the transport of heat is predominantly due to convective motions \citep{Spiegel1963,kraichnan1962}.
Indeed, these regimes have been observed in recent experiments on radiatively driven convection \citep{lepot2018, bouillaut2019}.
Hence, it is necessary to investigate the role of thermal boundary layers in turbulent convection to determine the asymptotic high Rayleigh number heat transport.

Motivated by the studies that used surface roughness to probe the boundary layers in turbulent shear flows, \citet{shen1996} studied turbulent thermal convection experimentally in a cell whose top and bottom surfaces were covered with pyramidal roughness elements of aspect ratio $2$, where the aspect ratio is the element width to height.
They observed that roughness led to the emission of a larger number of plumes compared to that in convection over smooth surfaces, and that when $Ra$ was above a certain threshold, $Nu$ increased by $20\%$ compared to its value for smooth surfaces.
However, the value of $\beta \approx 2/7$ was found to be the same as that for planar surfaces for the range of $Ra$ considered.
In later experiments, \citet{du1998, du2000} concluded similarly.
Several subsequent studies, however, report that roughness {\it does} lead to an increase in $\beta$ from its planar value \citep{roche2001, qiu2005, verzicco2006, tisserand2011, wei2014, salort2014, wagner2015, TSW15b, TSW17, zhu2017}.

The first study to use roughness to manipulate the interaction between the boundary layers and the outer region to attain the ultimate regime was that of \citet{roche2001} who studied convection experimentally in a cylindrical cell covered by V-shaped grooves on all sides.
They observed that when the thickness of the thermal boundary layers becomes smaller than the amplitude of roughness, $\beta$ attains a value of $0.51$ for $Ra = \left[2 \times 10^{12}, 5 \times 10^{13}\right]$.
Later, \citet{TSW15b, TSW17} used DNS in two-dimensions (2D) to systematically manipulate this interaction by varying the wavelength of sinusoidal upper and/or lower surfaces at a fixed amplitude.
They discovered the existence of an optimal wavelength at which $\beta$ is maximized, and that for wavelengths much smaller and much greater than the optimal wavelength $\beta$ attains its planar value.
They also found that $\beta = 0.483$ for the optimal wavelength when both top and bottom surfaces are corrugated \citep{TSW17}.
Their findings that roughness wavelength modulation leads to optimal heat transport and results in $\beta \approx 0.5$ for a certain wavelength were subsequently confirmed by experimental \citep{xie2017} and numerical \citep{zhu2017} studies, although it has been suggested that $\beta$ can decrease again at even higher Ra \citep{zhu2017}.
More recently, \citet{zhu2019} reported $Nu \sim Ra^{1/2}$ for $Ra = \left[10^8, 10^{11}\right]$ over corrugated surfaces with three characteristic length scales.

The central physical issue we are addressing here is as follows.
As emphasized above, the regimes of determining the exponent $\beta$ center around the interaction of the thermal boundary layers and the core flow.
As the Rayleigh number increases the thermal boundary layers thin.
Indeed, as first noted by \citet{niemela2006}, one can understand the results of \citet{roche2001} as a transition between a regime where the groove depth is less than the thermal boundary layer thickness to a regime where the groove depth is larger than the boundary thickness.
Thus, as emphasized by \citet{TSW17}, when a given experiment or simulation has a fixed roughness geometry, the boundary layer core flow interaction may evolve as the Rayleigh number increases.
It is for this reason that surfaces with a spectrum of roughness length scales are of interest.

Although we have considerable understanding of the effects of periodic corrugation on plume production and heat transport, it is still not clear {\it a priori} if these results could be used to describe the effects of fractal roughness.
Indeed, there have been far fewer studies on turbulent convection over multi-scale surfaces, the earliest being that of \citet{villermaux1998} who theoretically considered the effects of fractal surfaces with power-law distributed amplitudes.
\citet{villermaux1998} argued that given a regime where $Nu \sim Ra^{2/7}$ for smooth boundaries, the effective exponent increased from $2/7$ to $1/3$ with increasing degree of roughness. \citet{ciliberto1999} studied the effects of power-law distributed fractal surfaces on the heat transport experimentally and found larger $\beta$ values of $0.35$ and $0.45$ depending on the distribution of roughness amplitudes.
Those studies motivate our own.

In this work we consider the effects of one fractal boundary on the dynamics and bulk transport properties of turbulent Rayleigh-B\'enard convection and address the following questions:
(1) What are the effects of fractal surface roughness on the heat transport?
(2) How sensitive is the heat transport to the details of the roughness realization?
(3) Can one infer the characteristic length scale(s) of roughness from a study of its effects on the flow?
We do this using well resolved 2D numerical simulations using the Lattice Boltzmann Method. The choice of the domain and roughness properties is motivated by our aim to to understand the interactions between Arctic sea ice and the underlying ocean.

\section{Governing Equations}
The spatial domain in our study (in dimensional units) is $(x,z) \in [0,L]\times[0,h(x)]$ where $0 < h(x) \le H$ is the vertical height of the layer at horizontal position $x$.
We model thermal convection via the Oberbeck-Boussinesq equations \citep{Rayleigh1916, chandra2013}, non-dimensionalizing the system using the length scale $H$, the free-fall velocity scale $u_0 = \sqrt{g \, \alpha \, \Delta T \, H}$ where $g$ is acceleration of gravity, $\alpha$ is the coefficient of thermal expansion, $\Delta T$ is the temperature difference between the bottom and top boundaries, and the free-fall time scale $t_0 = H/u_0$. 

The equations and boundary conditions for the dimensionless velocity, temperature, and pressure fields $\boldsymbol{u}(\boldsymbol{x},t) = \left[u(\boldsymbol{x},t), w(\boldsymbol{x},t)\right]$, $T(\boldsymbol{x},t)$, and $p(\boldsymbol{x},t)$ are
\be
\nabla \cdot \boldsymbol{u} = 0,
\label{eqn:mass}
\ee 
\be
\frac{\partial \boldsymbol{u}}{\partial t} + \boldsymbol{u} \cdot \nabla \boldsymbol{u} = -\nabla p + T \, \boldsymbol{k} + \sqrt{\frac{Pr}{Ra}} \, \nabla^2 \boldsymbol{u},
\label{eqn:momentum}
\ee
\be
\frac{\partial T}{\partial t} + \boldsymbol{u} \cdot \nabla T = \sqrt{\frac{1}{Ra \, Pr}} \, \nabla^2 T,
\label{eqn:heat}
\ee
\be
\boldsymbol{u} = 0 \hspace{0.2cm} \text{and} \hspace{0.2cm} T = 1 \hspace{0.2cm} \text{at} \hspace{0.2cm} z = 0,
\ee
\be
\boldsymbol{u} = 0 \hspace{0.2cm} \text{and} \hspace{0.2cm} T = 0 \hspace{0.2cm} \text{at} \hspace{0.2cm} z = h(x).
\ee
The Rayleigh number $Ra = \alpha \, g \, \Delta T \, H^3/\kappa \, \nu$, where $\nu$ is the kinematic viscosity and $\kappa$ is the thermal diffusivity of the fluid, the Prandtl number $Pr = \nu/\kappa$. The fractal boundaries are such that $0.9 \le h(x) \le 1$.
All variables are periodic in the horizontal $x$ direction, and the aspect ratio of the domain is $\Gamma = L/H$.

The bulk heat transport is measured by the Nusselt number,
\be
Nu = \frac{\left<\overline{w^* \, T^*}\right> - \kappa \, \left<\frac{\partial \overline{T}^*}{\partial z^*}\right>}{\kappa \, \Delta T/H}, 
\label{eqn:Nu}
\ee
evaluated across horizontal layers in the cell.
Here the superscript $*$ indicates the variable is dimensional, and $\overline{(\cdot)}$ and $\left<\cdot\right>$ indicate horizontal and time averages, respectively.
We compute $Nu$ at eight different heights in the cell and report the average value over these locations.

We use the Lattice Boltzmann Method \citep{benzi1992, chen1998, succi2001} to solve the governing equations numerically.
The principal reason for this choice of numerical method is the ease with which one can impose the boundary conditions for the velocity and temperature fields on complicated domains \citep{succi2001}.
The code used here has been tested extensively in \citet{TSW15a} for different fluid flow problems \citep{clever1974,lipps1976, simakin1984} and was previously used to study turbulent convection over planar and corrugated (i.e., smooth but non-flat) upper and lower boundaries \citep{TSW15b, TSW17}.

We performed extensive checks on spatio-temporal convergence with the fractal boundaries used in the study. The spatial resolutions used in our simulations were such that the boundary layer was resolved with at least $8$ grid points and the Kolmogorov length scale was resolved with at least $2$ grid points everywhere in the domain. The simulations were run for sufficiently long times to attain a stationary state and the statistics were collected over the last $200$ time units, except for the $Ra = 10^{10}$ cases where the statistics were collected over the last $100$ time units. Details of these tests are provided in Appendix A. 

We should note here that $H$ is one of the many choices for the characteristic length scale for this geometry. However, this choice would only affect the pre-factor in power law scalings for Nusselt and Reynolds numbers with Rayleigh number, and not the exponent. (See Appendix B.) 

\section{Roughness Profiles}
Following \citet{rothrock1980}, we consider upper boundary functions $h(x)$ to be ``rough'' when they are continuous but not differentiable.
The increments in $h(x)$ are given by the H\"older condition,
\be
\lim_{\Delta x \to 0} \, \frac{\left|h(x+ \Delta x) - h(x)\right|}{(\Delta x)^{\gamma}} = C,
\label{eqn:lipschitz}
\ee
where $C$ is an ${\cal O}(1)$ constant and $0 < \gamma \le 1$ is the H\"older exponent.
Functions are Lipschitz continuous with a bounded derivative only when $\gamma = 1$.
The power spectral density (PSD) of $h(x)$ for all non-zero wavenumbers $k$ decays as $\sim k^{p}$, where $p = -2 \gamma - 1$ \citep{rothrock1980}.
This characteristic decay of the PSD is a common feature shared by many natural and artificial surfaces, and thus can be used to classify different classes of rough surfaces \citep{sayles1978, rothrock1980}.

To generate roughness profiles for the upper surface with the desired spectral properties for our simulations, we use the so-called truncated Steinhaus series \citep{rothrock1980};
\be
h(x) = h_0 + A \, \sum_{k=1}^{\mathcal{K}} \, \left(-p-1\right)^{1/2} \, k^{p/2} \, \cos(k \, x + \phi_k),
\label{eqn:steinhaus}
\ee
where $\mathcal{K}$ is the maximum wavenumber and $\phi_k$ are independent random variables uniformly distributed in $\left[0, 2\pi\right]$.
It is clear that the PSD of $h(x)$ in (\ref{eqn:steinhaus}) scales as $\sim k^p$ up to the cutoff wavenumber $\mathcal{K}$.

\begin{figure}
\begin{centering}
\includegraphics[trim = 0 0 0 0, width = 1\linewidth]{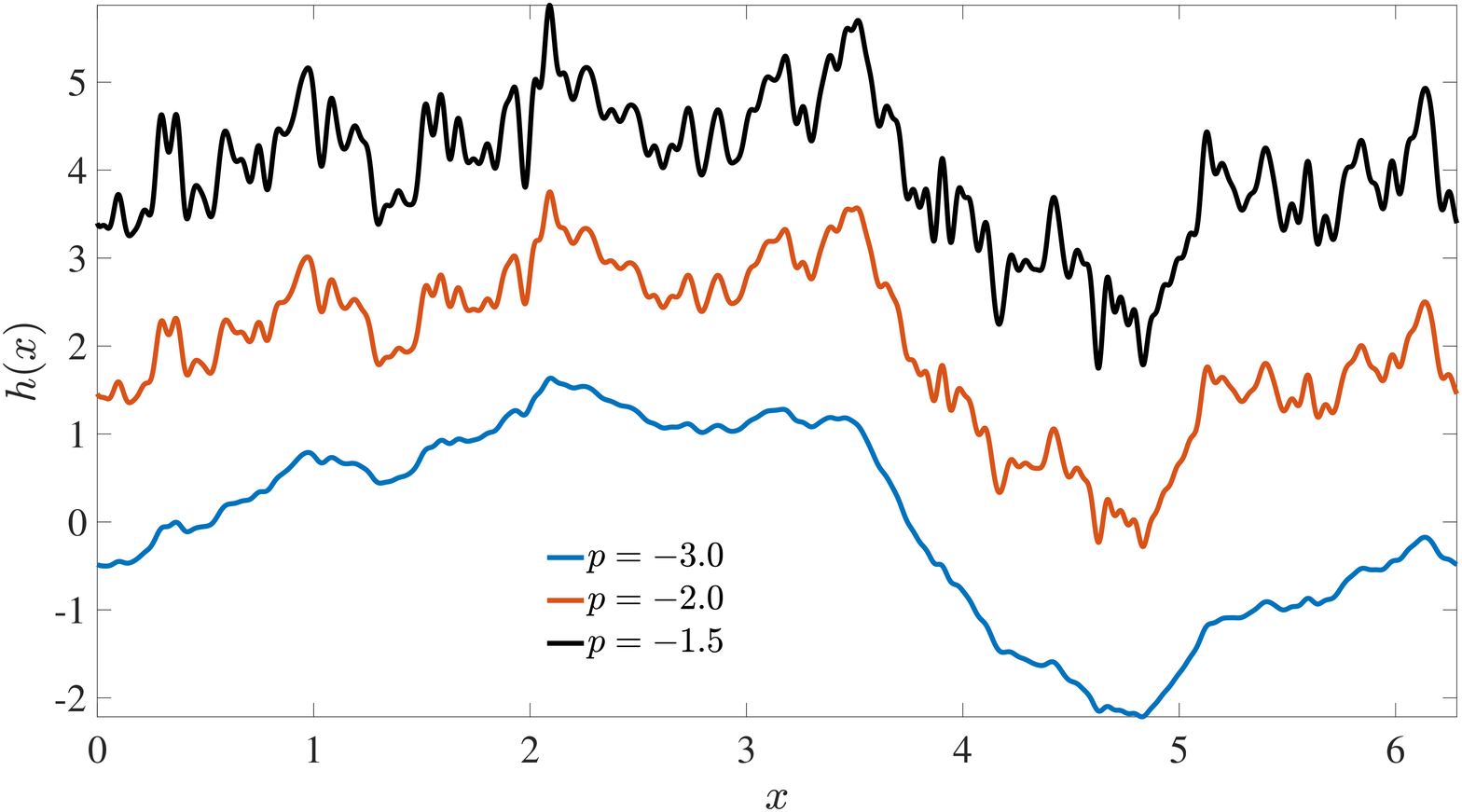} 
\caption{Functions used for the upper surface of the convecting domain generated using equation \ref{eqn:steinhaus} for different values of $p$ and for $\mathcal{K} = 100$.  The degree of roughness increases as the value of $p$ increases.  The curves are vertically displaced by $2$ units to improve their visibility.}
\label{fig:steinhaus}
\end{centering}
\end{figure}

Figure \ref{fig:steinhaus} shows roughness functions for different values of $p$ generated using equation (\ref{eqn:steinhaus}).
As $p$ is increased from $-3.0$ to $-1.5$, $h(x)$ becomes rough on smaller scales.
It is also intuitively clear from figure \ref{fig:steinhaus} that with increasing roughness, as $\mathcal{K} \rightarrow \infty$, $h(x)$ tends to be more space filling than a 1D curve but less space filling than a 2D surface.
Hence these curves are fractals in the limit $\mathcal{K} \rightarrow \infty$, with fractal or Hausdorff dimension $D_f = 2 - \gamma$ \citep{rothrock1980}.
We use equation (\ref{eqn:steinhaus}) to generate the rough upper surfaces $h(x)$ for the simulations. 
All the rough surfaces used in this study have $\mathcal{K} = 100$ and the values of $h_0$ and $A$ are chosen such that their maximum and minimum amplitudes, measured from the top of the cell, are $0\%$ and $10 \%$ of the depth of the cell, respectively. This implies that the upper portion of the fractal boundary coincides with the top flat surface. (See figure \ref{fig:temp_field}.)

From figure \ref{fig:steinhaus} it is clear that there is a distribution of amplitudes associated with the fractal curves; hence, it is not \emph{a priori} clear what choice of the characteristic length scale would be appropriate. The length scale chosen in our study is the depth of the cell $H$. We present a detailed discussion of this point in Appendix B and also show that a different choice of length scale simply leads to a uniform rescaling of $Nu$ and $Ra$ values for any given topography.

\section{Results}
The simulation results are for $Pr = 1$ and $\Gamma = 2$.
The simulations, except for $Ra = 10^{10}$, ran to at least $t = 330$ to allow adequate spin up, and in all cases the statistics were obtained for the last $200$ time units. The simulations for $Ra = 10^{10}$ were run for at least $t = 180$, and the statistics were collected over the last $100$ time units. See Appendix D for details.

\subsection{Temperature fields}
\begin{figure}
\centering

\begin{subfigure}
\centering
\includegraphics[trim = 0 0 0 0, clip, width = 1\linewidth]{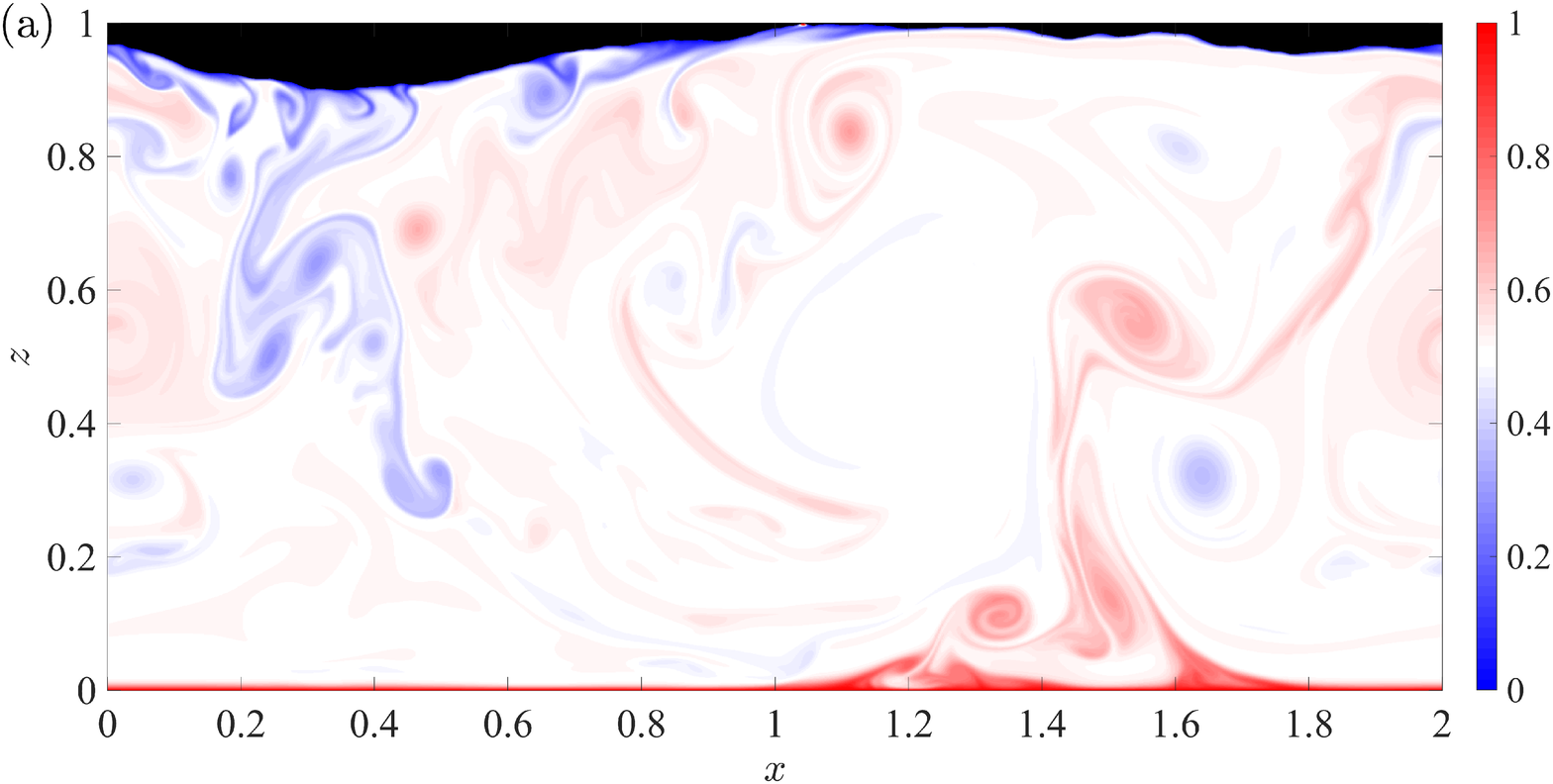}  
\end{subfigure}
    
\begin{subfigure}
\centering
\includegraphics[trim = 0 0 0 0, clip, width = 1\linewidth]{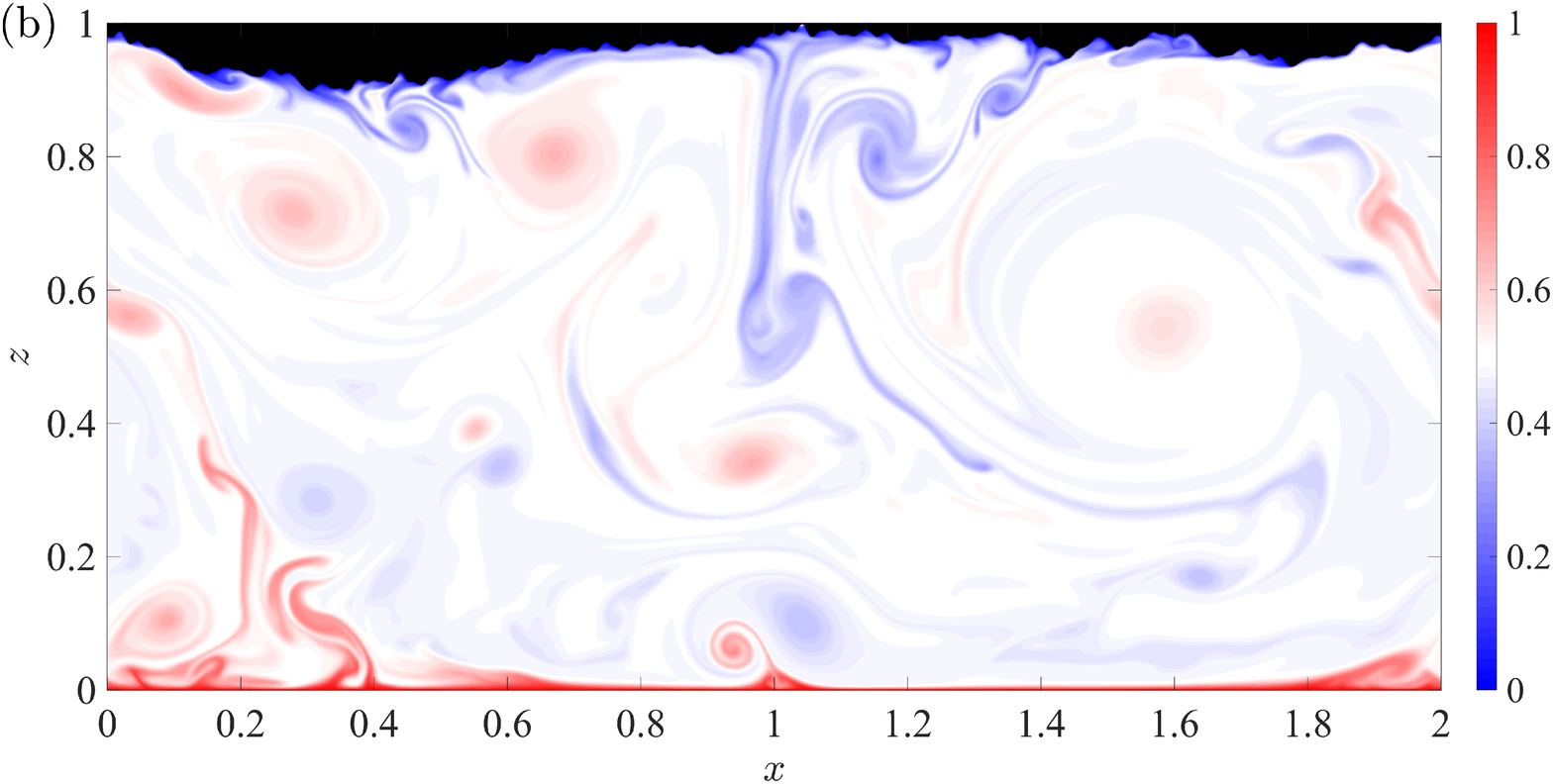} 
\end{subfigure}
          
\begin{subfigure}
\centering
\includegraphics[trim = 0 0 0 0, clip, width = 1\linewidth]{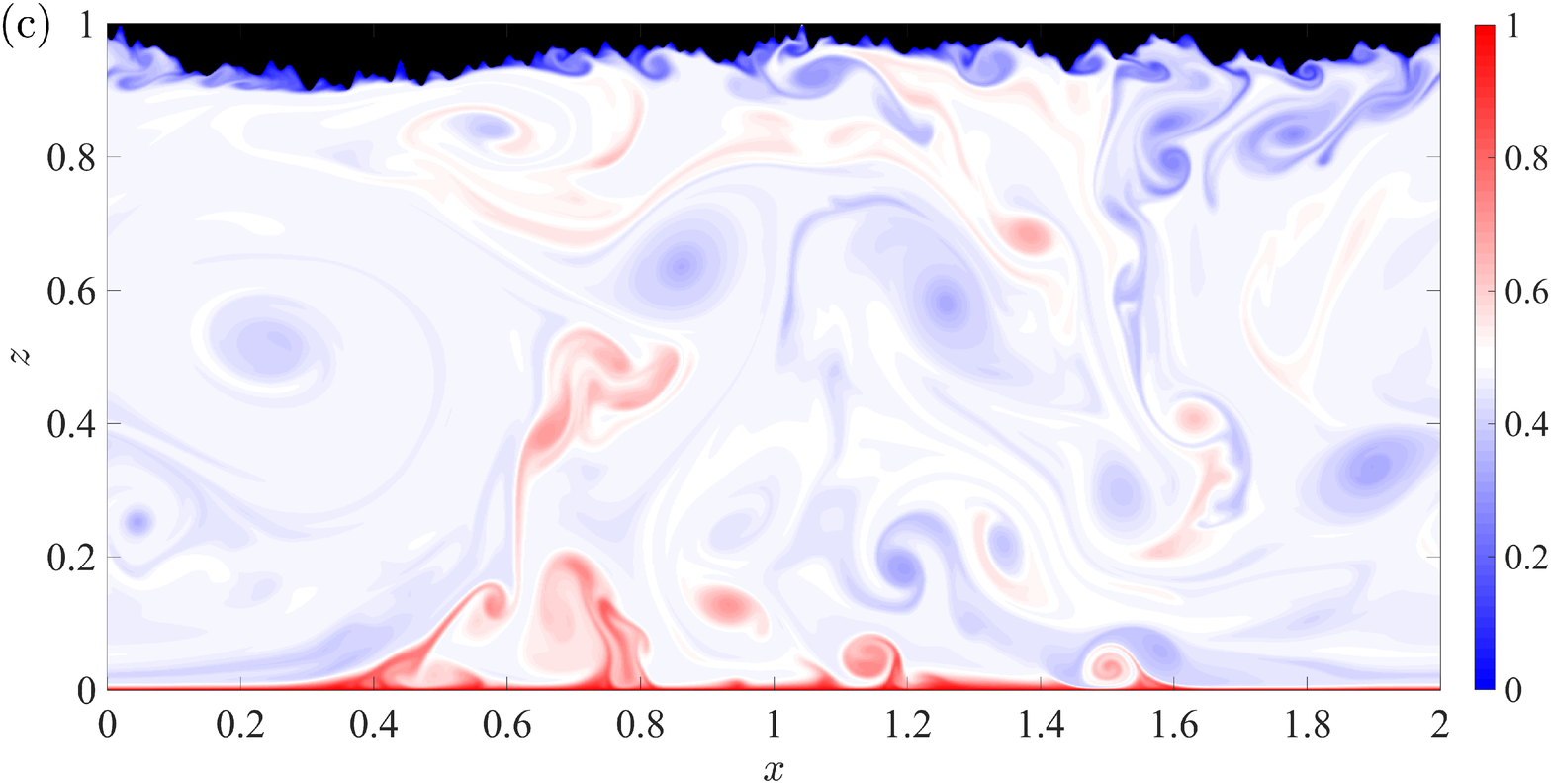}  
\end{subfigure}
          
\caption{Temperature fields at $t = 100$ for $Ra = 2.15 \times 10^9$ and (a) $p = -3.0$, (b) $p = -2.0$, and (c) $p = -1.5$.}     
\label{fig:temp_field}             
\end{figure}

Figures \ref{fig:temp_field} (a)--(c) show the snapshots of the temperature fields for $Ra = 2.15 \times 10^9$ and $p = -3$, $-2$, and $-1.5$, respectively.
Focusing on the region close to the rough upper surfaces, for $p = -3$ (which is comparatively smooth) plumes are emitted only from a fraction of the surface and the temperature field is qualitatively similar to that for convection over flat walls \citep[e.g.,][]{doering2009}. 
As seen in figures \ref{fig:temp_field}(b) and (c), however, as $p$ increases so too do the number of roughness elements triggering more plume generation: roughness enhances the coupling between the boundary layer and the core flow.
Moreover, as seen in the case of a periodically corrugated upper surface \citep{TSW15b}, the enhanced emission of cold plumes decreases the mean interior temperature relative to the planar surface case.

\subsection{Variation of heat flux with roughness properties} \label{sec:NuRa}
\begin{figure}
\centering

\begin{subfigure}
\centering
\includegraphics[trim = 0 0 0 0, clip, width = 1\linewidth]{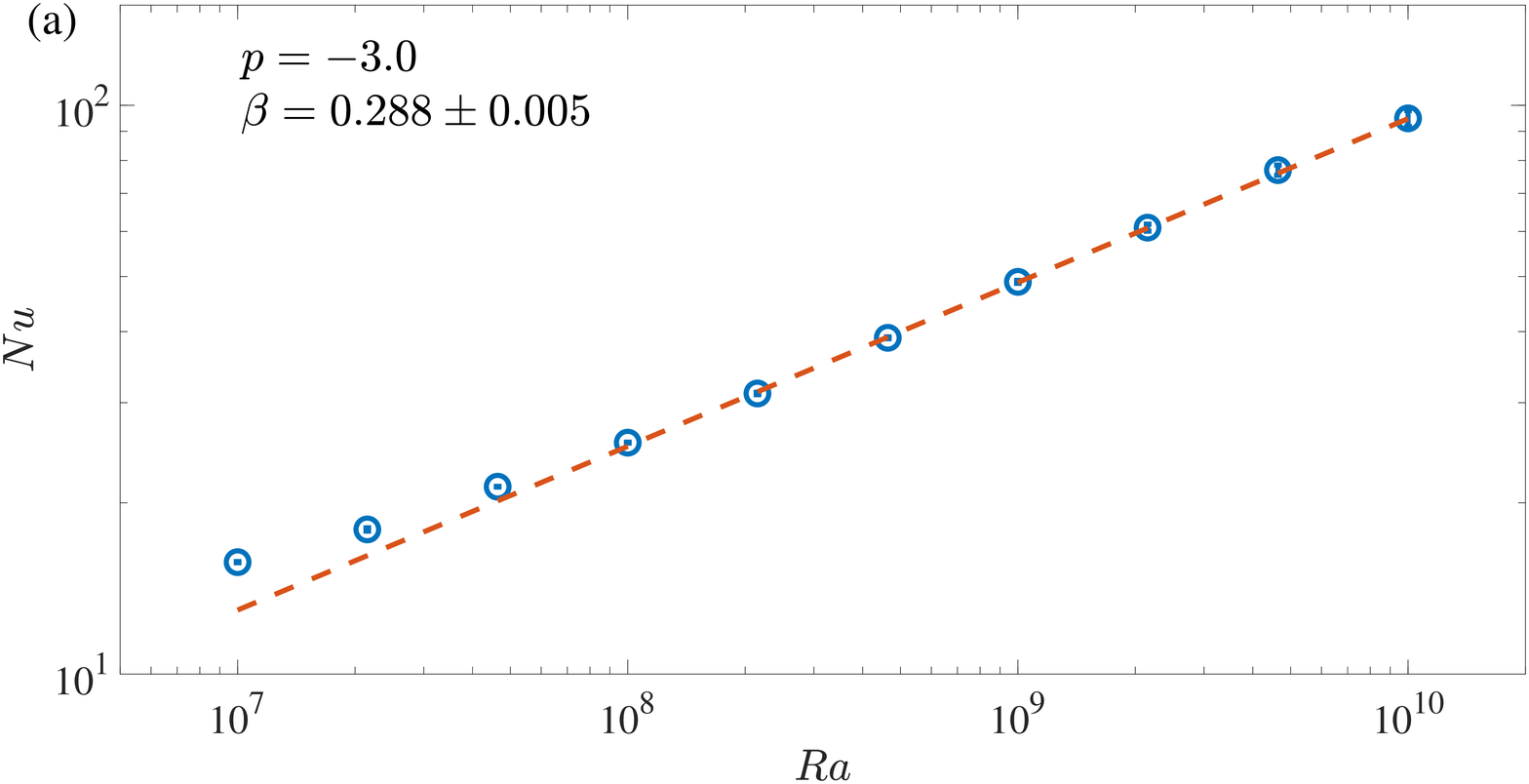}  
\end{subfigure}
    
\begin{subfigure}
\centering
\includegraphics[trim = 0 0 0 0, clip, width = 1\linewidth]{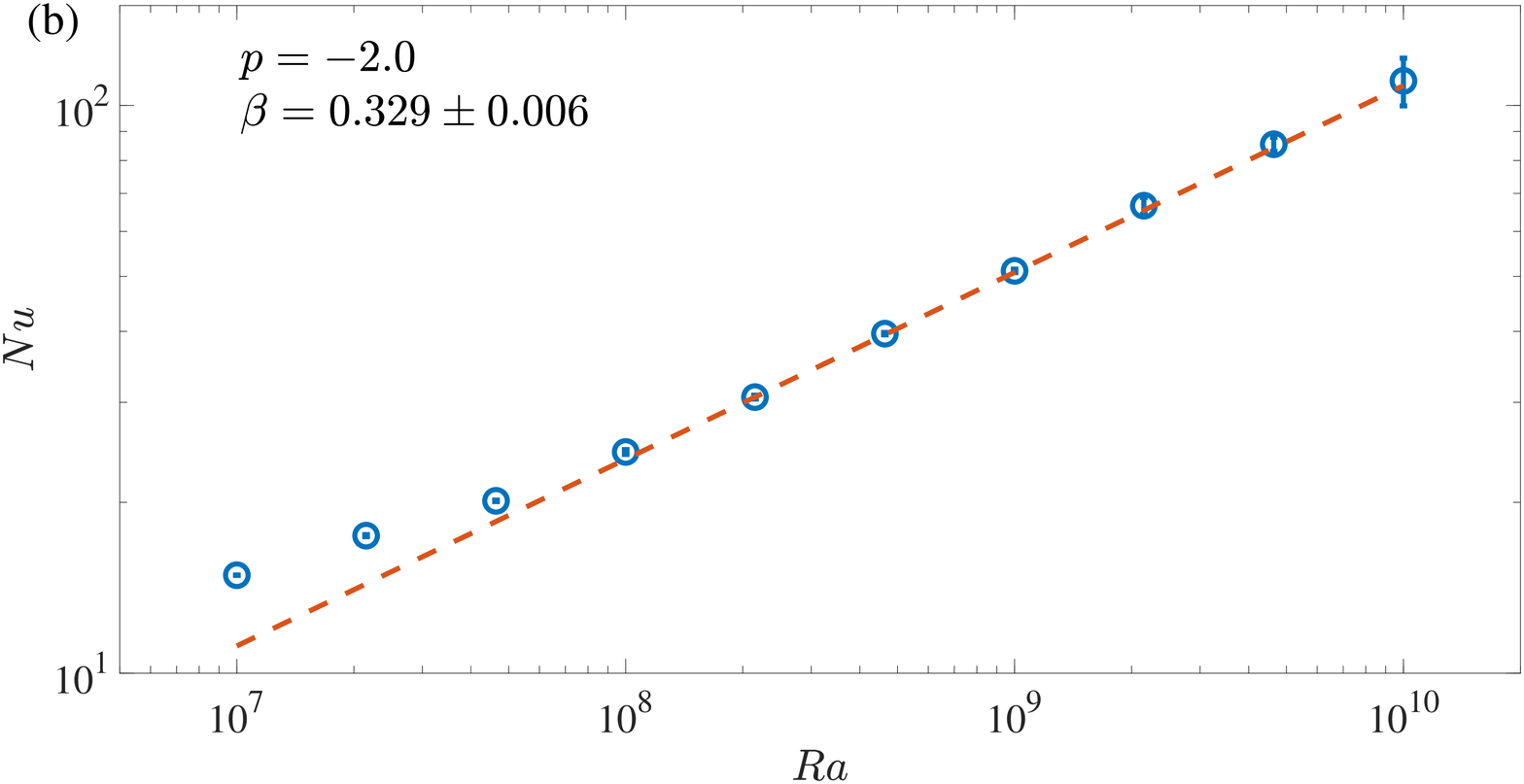} 
\end{subfigure}
          
\begin{subfigure}
\centering
\includegraphics[trim = 0 0 0 0, clip, width = 1\linewidth]{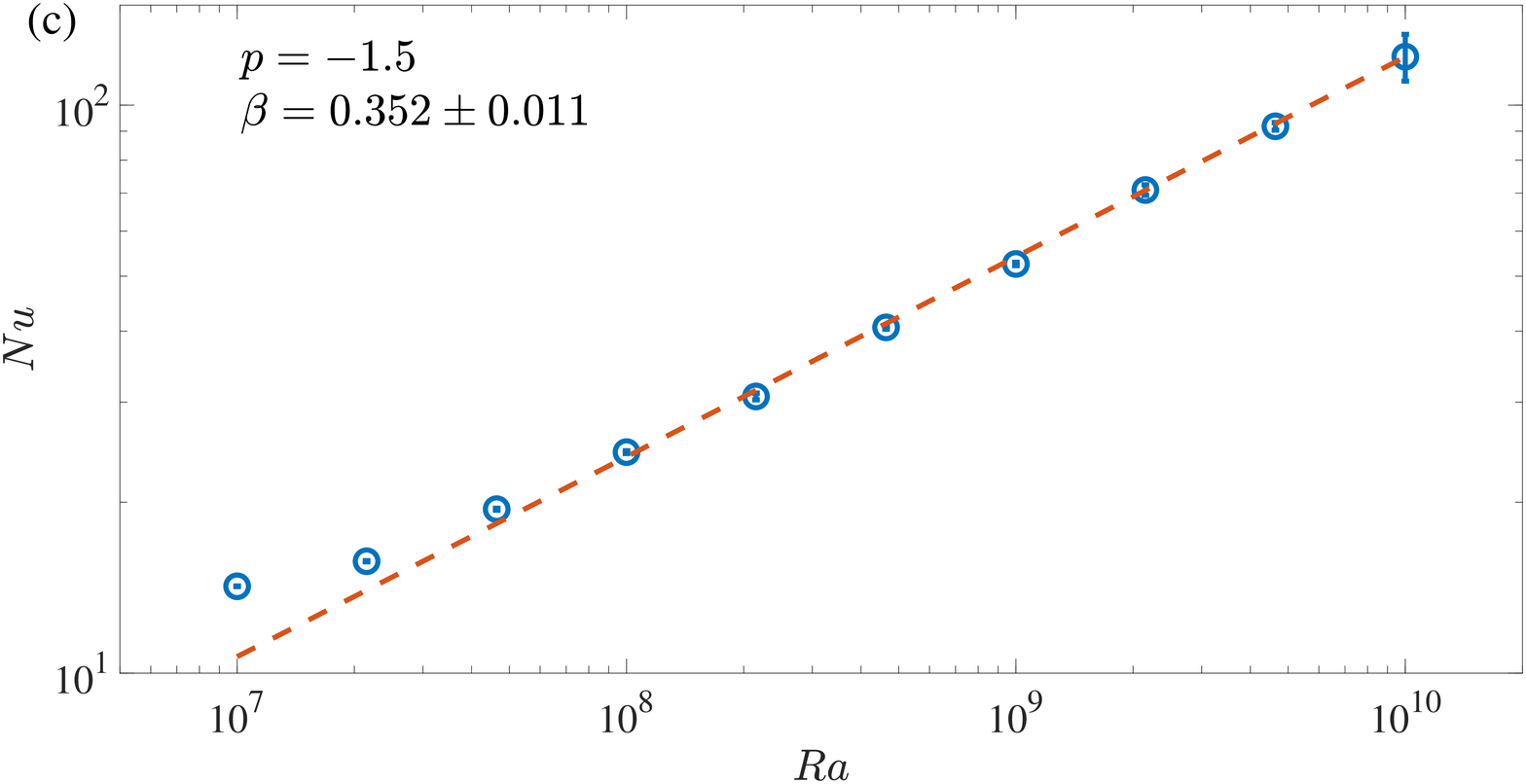}  
\end{subfigure}

\caption{$Nu(Ra)$ vs. $Ra \in \left[10^7, 10^{10}\right]$ and (a) $p = -3.0$, (b) $p = -2.0$, and (c) $p = -1.5$. Circles denote data from simulations and the dashed lines are the linear least-squares fits of $\log Nu$ to $\log Ra$ over the range $Ra \in \left[10^8, 10^{10}\right]$. The power laws are for the range $Ra \in \left[10^8, 10^{10}\right]$. For (a) $p=-3.0$, $Nu = 0.125 \times Ra^{0.288 \pm 0.005}$; (b) $p = -2.0$, $Nu = 0.055 \times Ra^{0.329 \pm 0.006}$; and (c) $p = -1.5$, $Nu = 0.037 \times Ra^{0.352 \pm 0.011}$. The error bar on each $Nu$ data point represents the standard deviation of the averaged $Nu$ calculated from eight different horizontal sections, and the uncertainties in $\beta$ are the $95\%$ confidence intervals.}
\label{fig:NuRa}             
\end{figure}

The $Nu(Ra)$ data are shown in figures \ref{fig:NuRa}(a)--(c) for $Ra \in \left[10^7, 10^{10}\right]$ and (a) $p = -3.0$, (b) $p = -2.0$, and (c) $p = -1.5$, respectively. In these figures, for a given value of $p$, the simulations for the whole $Ra$ range were performed using the same realization of the fractal boundary. The scaling fits (i.e., linear least squares of the logarithms) are for $Ra \in \left[10^8, 10^{10}\right]$ and are shown as dashed lines in these figures.
When $p$ increases from $-3.0$ to $-1.5$, the scaling exponent increases from $\beta = 0.288$ to $\beta = 0.352$. The power-law fit to the $Nu(Ra)$ data for $p=-3.0$ gives $Nu = 0.125 \times Ra^{0.288 \pm 0.005}$, which is remarkably close to $Nu = 0.138 \times Ra^{0.285}$ obtained for a similar $Ra$ range for flat boundaries in 2D \citep{doering2009}. This suggests that, for this range of $Ra$, the fractal boundary corresponding to $p=-3.0$ is hydrodynamically smooth with respect to heat transport, even though it lies at the border between smooth and rough boundaries \citep{rothrock1980}. For $p = -2.0$, the power law fit gives $Nu = 0.055 \times Ra^{0.329 \pm 0.006}$, which is surprisingly close to $Nu = (0.0525 \pm 0.006) \times Ra^{0.331 \pm 0.002}$ obtained for $Ra \in \left[10^{10}, 10^{15}\right]$ for flat boundaries in a slender cylinder with $\Gamma = 0.1$ \citep{iyer2020}. And, lastly, for $p = -1.5$ the power law is $Nu = 0.037 \times Ra^{0.352 \pm 0.011}$, which is remarkably close to $Nu = 0.034 \times Ra^{0.359}$ obtained for (scaled) wavelength, $\lambda = 0.154$ for a sinusoidally corrugated boundary in 2D over $Ra \in \left[4 \times 10^6, 2.5 \times 10^9\right]$ \citep{TSW15b}. In fact, \citet{TSW15b} found that $\lambda = 0.154$ was the optimal wavelength that maximized heat transport for their geometry. Evidently, heat transport increases with increasing degree of roughness i.e., with $p$. This is in qualitative agreement with the results of \citet{villermaux1998}, who also found that the scaling exponent increases with increasing roughness.  However, the increase in $\beta$ is due to a change in the dynamics brought about by the introduction of surface roughness, principally the increase in plume production \citep{verzicco2006, TSW15b, TSW17}, and not due to increased surface area. See Appendix C for details. We also note that the power-law fits to the whole range of $Ra$ give: (a) $p = -3.0$: $Nu = 0.197 \times Ra^{0.267 \pm 0.015}$; (b) $p = -2.0$: $Nu = 0.111 \times Ra^{0.30 \pm 0.02}$; and (c) $p = -1.5$: $Nu = 0.069 \times Ra^{0.321 \pm 0.023}$.

\begin{figure}
\begin{centering}
\includegraphics[trim = 0 0 0 0, width = 1\linewidth]{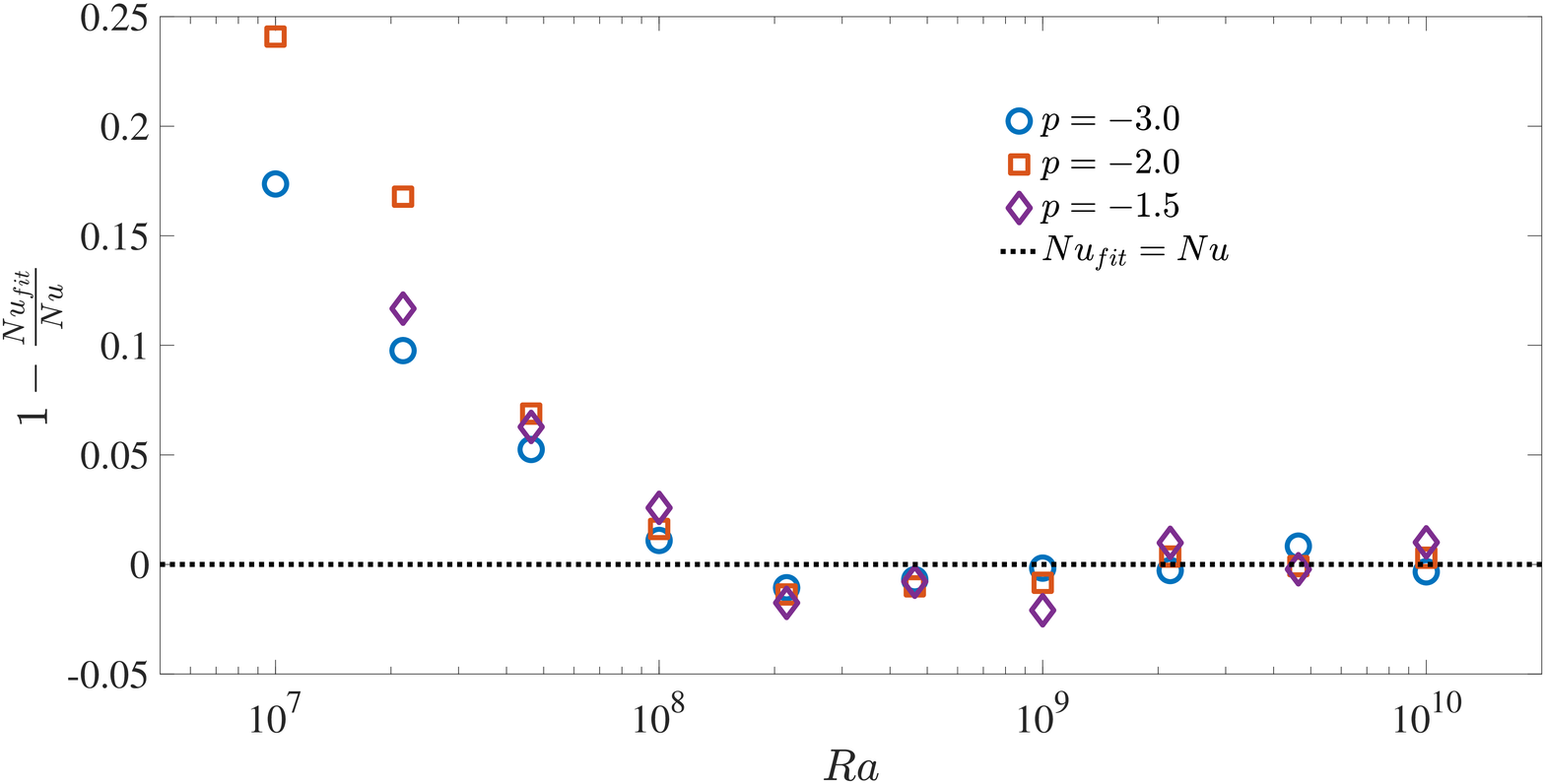} 
\caption{Residuals of the power-law fits shown in figure \ref{fig:NuRa} for $p = -3.0, -2.0$ and $-1.5$. Here, $Nu_{fit}$ are the values of the Nusselt number obtained from the power-law fits for the range $Ra \in \left[10^8, 10^{10}\right]$.}
\label{fig:residuals}
\end{centering}
\end{figure}

The goodness of power-law fits can be tested by computing the residuals of the actual data from the fit data and examining them for any systematic curvature. These are shown in figure \ref{fig:residuals}.
It is apparent that the residuals do not exhibit any curvature for $Ra \in \left[10^8, 10^{10}\right]$; hence, it can be concluded that the power laws describe the data well and more general fits are not required.

We can estimate an effective hydrodynamic length scale for the roughness amplitude by examining where the $Nu(Ra)$ curves for different values of $p$ intersect.
When boundary variations are present, the flow is not influenced by the roughness until the boundary layers become smaller than the amplitude of roughness \citep{roche2001}.
Once this is achieved, as $Ra$ increases further the direct effects of roughness are associated with the increased number of plumes produced and the concomitant augmentation in $Nu$ \citep{roche2001}.
We use similar ideas to estimate the effective roughness of the fractal boundaries.
As seen in figures \ref{fig:steinhaus} -- \ref{fig:NuRa}, the additional roughness structure introduced as $p$ is increased is associated with the increase in $\beta$.
If one takes the $Nu-Ra$ curve for $p = -3$ as the benchmark case, then the intersection of this curve with that for a larger value of $p$ gives the value of the effective amplitude at which the transition to enhanced heat transport occurs.
The choice of this reference is because $p=-3$ corresponds to $\gamma = 1$, representing the border between ``smooth'' and ``rough'' surfaces \citep{rothrock1980}.
Thus the effects of any additional roughness (see figure \ref{fig:steinhaus}) can be conveniently studied with respect to the surface for $p=-3$.
This transition happens at $Ra \approx 2.15 \times 10^8$, and the value of $Nu$ at this point is $\approx 31$ (see the $Nu$ values for the fourth realizations ($r = 4$) in tables \ref{tab:p3_NuRa} -- \ref{tab:p32_NuRa}).
Hence, the transition occurs when the effective amplitude of roughness $h_f$ over the surfaces with $p = -2$ and $p = -1.5$ first exceeds the boundary layer thickness $\delta_T$ for the curve with $p = -3$, so that the roughness elements protrude outside of the boundary layer and interact with the interior of the flow.
Using the planar-wall estimate of $Nu$, we estimate the crossover scaling
\be
\delta_T = h_f \approx \frac{1}{2 \, Nu} = 0.016.
\label{eqn:amplitude}
\ee
Thus the effective amplitude of the roughness for surfaces with $p=-2$ and $p=-1.5$ is about $2\%$ of the depth of the cell.

\subsection{Sensitivity of $Nu$ to details of roughness realization} \label{sec:Nu_sensitivity}
\begin{figure}
\centering

\begin{subfigure}
\centering
\includegraphics[trim = 0 0 0 0, clip, width = 1\linewidth]{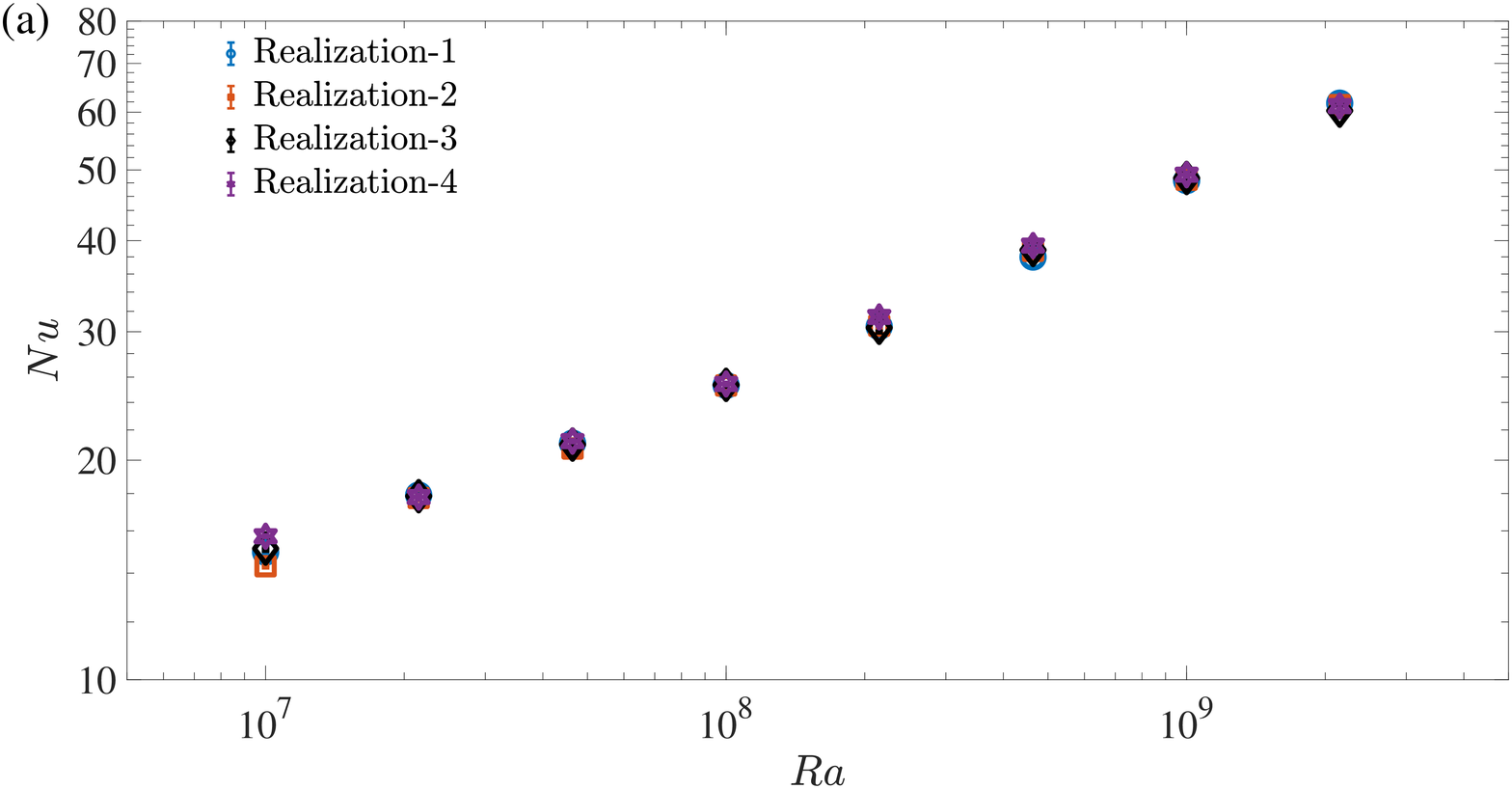}  
\end{subfigure}
    
\begin{subfigure}
\centering
\includegraphics[trim = 0 0 0 0, clip, width = 1\linewidth]{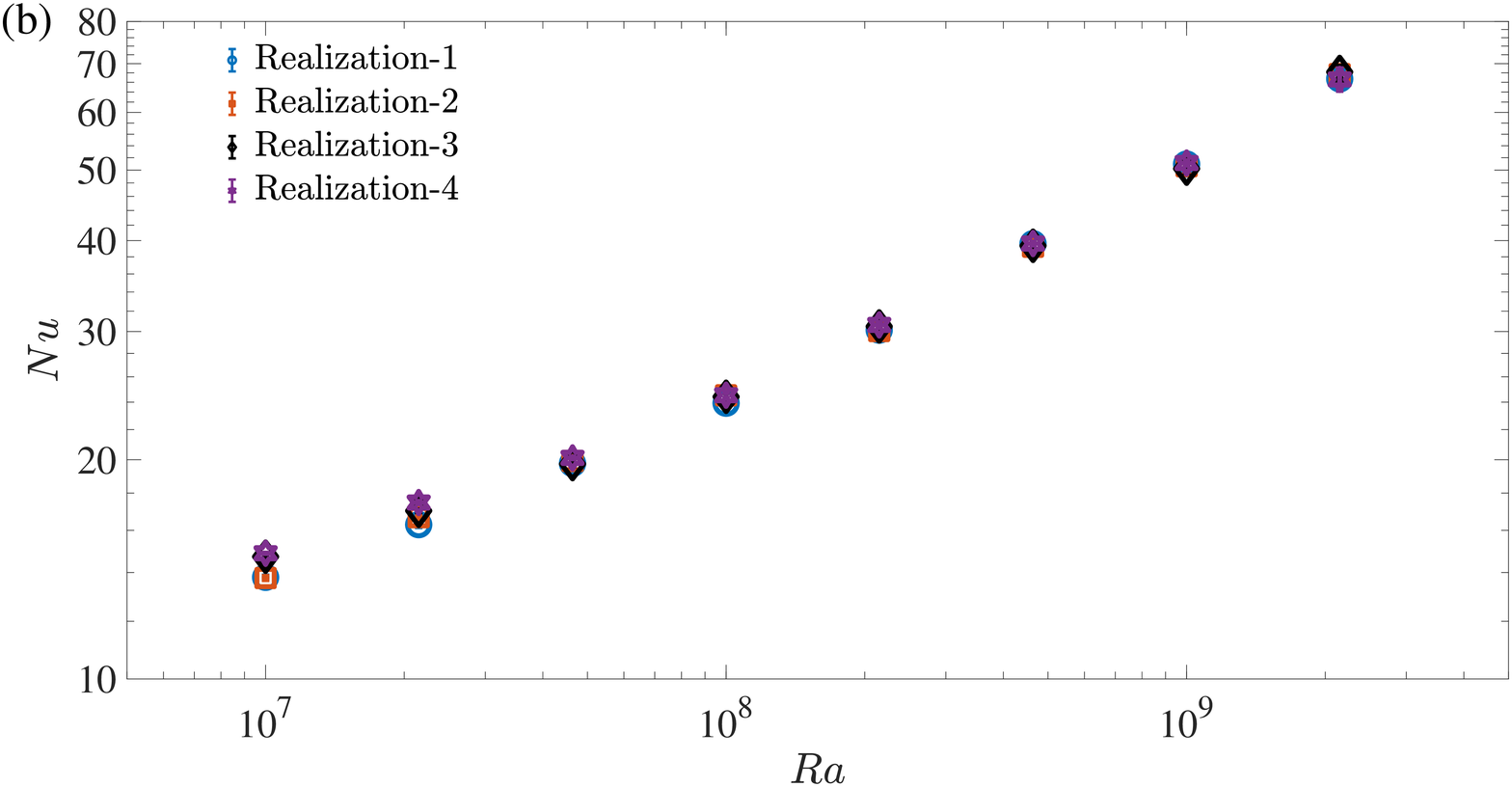} 
\end{subfigure}
          
\begin{subfigure}
\centering
\includegraphics[trim = 0 0 0 0, clip, width = 1\linewidth]{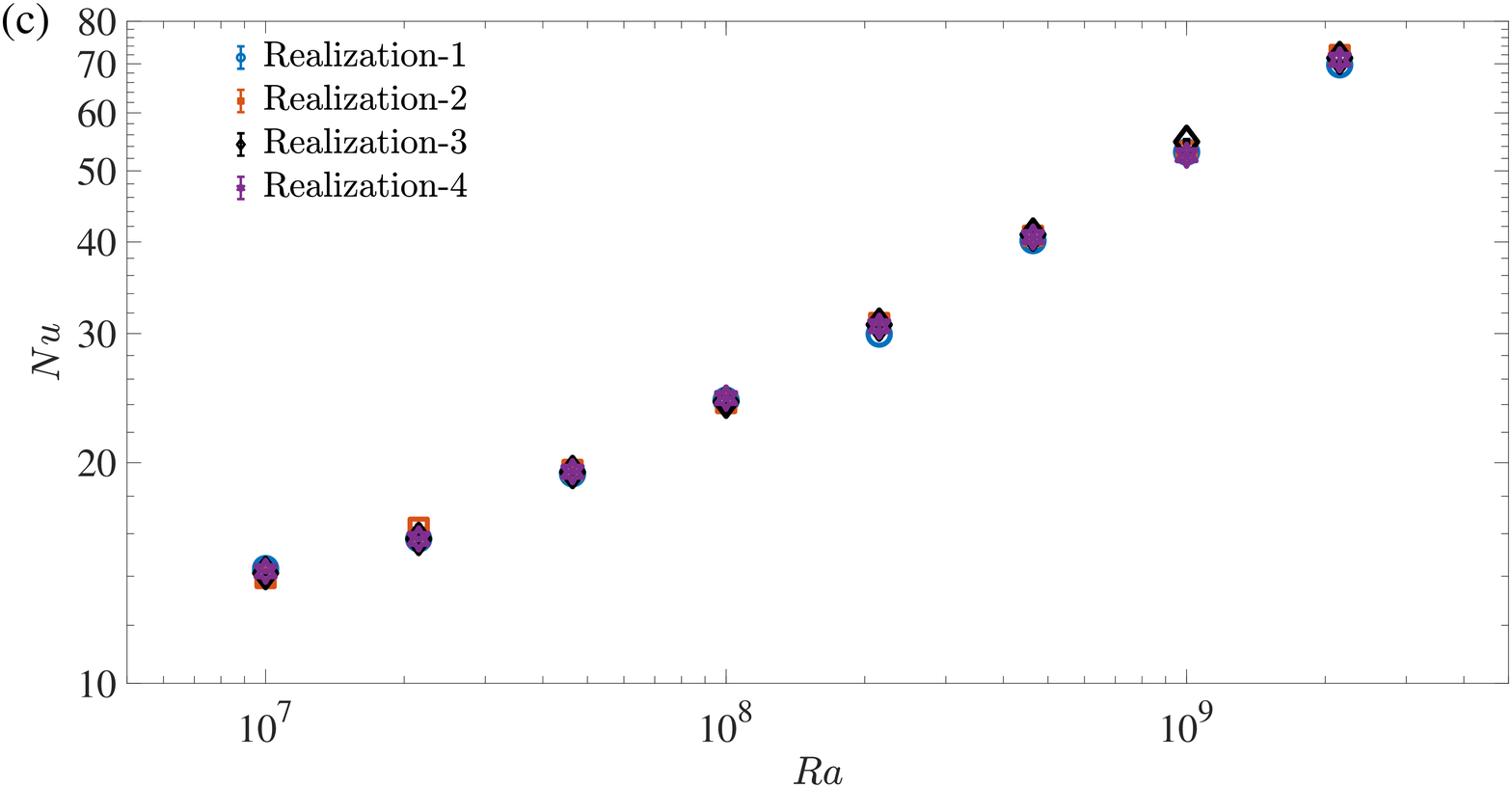}  
\end{subfigure}

\caption{$Nu(Ra)$ data for four different realizations for (a) $p = -3.0$, (b) $p = -2.0$, and (c) $p = -1.5$. The error bar on each $Nu$ data point represents the standard deviation of the averaged $Nu$ calculated from eight different horizontal sections.}
\label{fig:NuRa_realization}             
\end{figure}

To investigate the effects of a given roughness realization on heat transport, we computed $Nu(Ra)$ for four different realizations for each value of $p$. To generate each realization for a fixed $p$, we have used different values of $\phi_k$. However, the first realization for all $p$'s have the same set of $\phi_k$. Similarly for second, third, and fourth realizations. The $Nu(Ra)$ curves from these simulations are shown in figure \ref{fig:NuRa_realization}.

It is seen from figures \ref{fig:NuRa_realization}(a) -- \ref{fig:NuRa_realization}(c) that for a fixed $p$, the $Nu$ depends primarily on the $Ra$ with very little dependence on the realization. Hence, to a good approximation, the heat transport for the fractal surfaces used here depends only the latter's statistical properties, i.e., $p$, and in turn on $D_f$. Hence, this suggests that the scaling exponent $\beta$ depends uniquely on $p$.

Furthermore, to compute higher order moments, we have run simulations for $Ra = 10^8$ and $t = 875$ for all the roughness realizations. The maximum variations in the means of $Nu(t)$ measured at $z=0$ between ensemble members for for $p = -3.0, -2.0$, and $-1.5$ are $3.3\%$, $1\%$, and $0.2\%$, respectively. Similarly, the maximum variations in the standard deviations for $p = -3, -2$, and $-1.5$ are $5.4\%$, $16\%$, and $9.1\%$, respectively. The variations in the higher-order moments (skewness and kurtosis) are relatively larger. This suggests that the mean of $Nu(t)$ is less sensitive to the details of the roughness than its higher-order moments.

\subsection{Reynolds number and its sensitivity to the details of the roughness realization}
In addition to considering the bulk heat transport, we also studied the behavior of the bulk Reynolds number ($Re$) with $Ra$ and $p$ to further characterize the response of the flow.
The Reynolds number is
\be
Re = \frac{U_0 \, H}{\nu},
\ee
where $U_0$ is a velocity scale, the choice of which is not unique.
Previous studies over smooth \citep{qiu2001, niemela2001, sun2005, niemela2006} and regular rough surfaces \citep{wei2014} have either constructed $U_0$ based on the depth of the cell and the dominant frequency of oscillations of the large-scale circulation, or used a root-mean-squared (RMS) velocity deduced from single-point measurements.
We take $U_0 = U_{rms}$, where $U_{rms}$ is the bulk averaged RMS velocity computed over all the nodes in the domain.

\begin{figure}
\begin{centering}
\includegraphics[trim = 0 0 0 0, width = 0.9\linewidth]{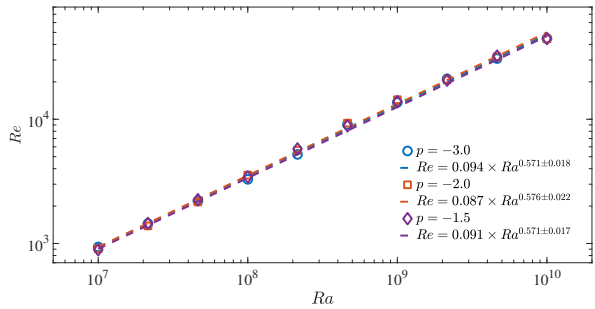} 
\caption{$Re(Ra)$ vs. $Ra = \left[10^7, 10^{10}\right]$ and $p = -3.0$, $p = -2.0$, and $p = -1.5$. Symbols denote data from simulations and the dashed lines are the linear least-squares fits of $\log Re$ to $\log Ra$ for the whole $Ra$ range. For (a) $p=-3.0$, $Re = 0.094 \times Ra^{0.571 \pm 0.018}$; (b) $p = -2.0$, $Re = 0.087 \times Ra^{0.576 \pm 0.022}$; and (c) $p = -1.5$, $Nu = 0.091 \times Ra^{0.571 \pm 0.017}$. The uncertainties in the values of $\xi$ are the $95\%$ confidence intervals.}
\label{fig:ReRa}
\end{centering}
\end{figure}

Figure \ref{fig:ReRa} shows $Re(Ra)$ data along with power-law fits $Re \sim Ra^{\xi}$ for the three different $p$.
Unlike $\beta$, the exponent $\xi$ characterizes scaling behavior of $Re$ over three full decades of $Ra$.
Moreover, $Re(Ra)$ is substantially less sensitive to details of the roughness: $\xi \approx 0.57$ for all three values of $p$ and the prefactor variation among the three values of $p$ is less than 8\%.
This suggests that the strength of the velocity variations in the cell is set by the large scale properties of the boundary profile that are present for the smooth surface with $p=-3$, and that smaller scale roughness does not appreciably affect $\xi$. Recent observation of $Re \sim Ra^{0.617}$ scaling for turbulent convection over flat boundaries in 2D \citep{wan2020} is consistent with this suggestion.

Note that $\xi = 0.5$ corresponds to the (dimensional) RMS fluid speed being proportional to the free-fall velocity across the cell, $u_0 = \sqrt{g \alpha \Delta T H}$.  Because the boundary temperatures are fixed, $g \alpha \Delta T$ is the maximal buoyancy acceleration of any fluid element, so suitably conspiratorial flow configurations would be required to sustain $\xi > 0.5$ as $Ra \rightarrow \infty$. Such is the case in coherent steady (albeit unstable) convection between stress-free boundaries where $\xi \rightarrow 2/3$ as $Ra \rightarrow \infty$ \citep{chini2009}.

\begin{figure}
\begin{centering}
\includegraphics[trim = 0 0 0 0, width = \linewidth]{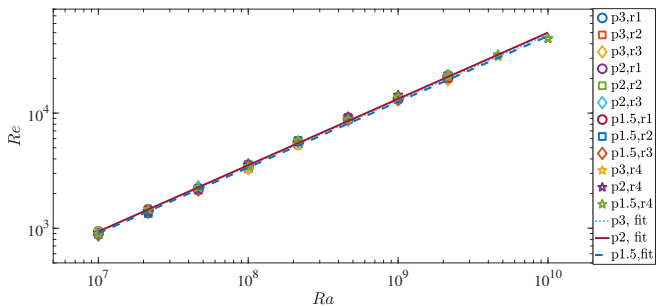} 
\caption{$Re(Ra)$ data for all $p$ and the different realizations. The power-law fits shown here are the ones reported in figure \ref{fig:ReRa} for $Ra \in \left[10^7, 10^{10}\right]$.}
\label{fig:Re_realization}
\end{centering}
\end{figure}

In order to understand the effects of details of roughness realization on the variation of $Re(Ra)$, we performed an analysis similar to that reported for $Nu(Ra)$ data. Figure \ref{fig:Re_realization} shows $Re(Ra)$ data for all $p$ and the different realizations. The realizations used here are the same as those used for obtaining the $Nu(Ra)$ data.
Figure \ref{fig:Re_realization} clearly suggests that $Re(Ra)$ is independent of the details of the roughness realizations and the value of $p$ itself. This is further supported by the fact that the power-law fits to $Re(Ra)$ data for all four different realizations for each $p$ and $Ra \in \left[10^7,2.15\times 10^9\right]$ give: (a) $p=-3.0$: $Re = 0.073 \times Ra^{0.584}$; (b) $p = -2.0$: $Re = 0.069 \times Ra^{0.588}$; and (c) $p = -1.5$: $Re = 0.068 \times Ra^{0.589}$. Hence, unlike $\beta$, $\xi$ is independent of the roughness geometries used in this study.

\section{Conclusions}

We have systematically studied turbulent thermal convection in domains with a fractal upper boundary for $Ra \in \left[10^7, 10^{10}\right]$ in two-dimensions using the Lattice Boltzmann Method.
The fractal nature of the boundaries are characterized by their spectral exponent $p=2 D_f -5$ representing the degree of roughness, where $D_f$ is the Hausdorff dimension of the boundary function.
Simulations with roughness exponents $p = -3.0, -2.0$ and $-1.5$ revealed the following:
\begin{enumerate}

\item With increasing roughness, the fractal boundaries provide an increasing number of sites for the generation of plumes.
Hence, at fixed $Ra$ the plume production increases with increasing $p$.

\item The $Nu \sim Ra^{\beta}$ power-law fit exponent $\beta$ for the range $Ra \in \left[10^8, 10^{10}\right]$ increased from $0.288$ to $0.352$ as $p$ increased from $-3.0$ to $-1.5$. Heat transport increased with roughness for larger $Ra$, in qualitative agreement with \citet{villermaux1998}. This increase in $\beta$ is due to a change in the dynamics that results from the enhanced interactions between the rough boundary and the inner flow, through the increase in plume production \citep{verzicco2006, TSW15b, TSW17}. The increased surface area over the boundaries also increases heat transport, but only by increasing the pre-factor in the power law and not $\beta$. (See Appendix C.)

\item The fractal surfaces used in the experiments of \citet{ciliberto1999} were built with glass spheres such that the amplitudes were power-law distributed, i.e.,  $P(h) \sim h^{\Lambda}$. They reported $\beta = 0.35$ for $\Lambda = -2$ and $\beta = 0.45$ for $\Lambda = -1$. Hence, $\beta$ increased with increasing $\Lambda$, which represents the degree of roughness. Hence, our findings are qualitatively consistent with the results of \citet{ciliberto1999}.

\item The following observations can be made based on the analysis of $Nu-Ra$ data: (a) the fractal boundary for $p=-3.0$ is hydrodynamically smooth for heat transfer, as both the exponent and prefactor in the $Nu-Ra$ power law are approximately those observed in convection over flat boundaries \citep{doering2009}; (b) both the prefactor and exponent of the $Nu-Ra$ power law for $p = -2.0$ correspond surprisingly well, albeit perhaps only fortuitously, to those reported for convection over smooth surfaces in a small aspect ratio 3D cylindrical geometry \citep{iyer2020}; and (c) the $Nu-Ra$ power law for $p=-1.5$ is remarkably close to the one obtained for the optimal wavelength of a corrugated sinusoidal boundary that maximizes heat transport \citep{TSW15b}.

\item Using the roughness curve for $p = -3$ as the reference profile, we estimated the effective amplitudes of the roughness curves for $p=-2$ and $-1.5$ that lead to increased heat transport. This is about $2\%$ of the depth of the cell for $p = -2$ and $p = -1.5$.

\item The averaged $Nu$ values for a fixed $p$ depend primarily on $Ra$ and have a very weak dependence on the roughness realization. However, the higher-order moments are more sensitive to the details of roughness realizations.

\item The Reynolds numbers based on the RMS velocity computed over all fluid nodes scaled as $Re \sim Ra^{\xi}$, with $\xi \approx 0.57$, for all three values of $p$ studied here. Perhaps surprisingly, the bulk intensity of the flow was substantially less sensitive to small-scale details in the roughness profiles than the heat transfer.

\item Like the averaged $Nu$ values, the averaged $Re$ values for a fixed $p$ depend primarily on $Ra$, and have a weak dependence on the roughness realization.

\item To a good approximation, the exponent $\beta$ is solely a function of $p$ (and in turn $D_f$), and $\xi$ is independent of $p$.

\end{enumerate}

These simulations demonstrate the feasibility of studying turbulent flows over fractal walls using numerical simulations.
Importantly, they provide a framework to study heat transport in high $Ra$ convection that can reveal the influence of interactions between the boundary layers and core flow.
Namely, we know that such interactions are important for the $Nu(Ra)$ behavior and that as $Ra$ increases, boundary layers thin and so too will the size of roughness elements that trigger plume production.
For a given fractal surface, only a fraction of the roughness elements are driving boundary layer instability and that fraction changes with $Ra$.
Therefore fractal surfaces that enhance plume production and heat transport must also optimize the fraction of the ``active'' surface roughness elements.
However, although a fractal surface reveals finer details with increasing resolution, all numerical simulations are ultimately limited by finite resolution so there will always be details of the surface that the flow would not be able to sense.
This leads naturally to the question of how one can represent the effects these unresolved details of roughness on the turbulent flows, a perennial conundrum in all manner of flows adjacent to surfaces.

\section*{Acknowledgements}
The authors acknowledge the support of the University of Oxford, NORDITA, and Yale University.
S.T. acknowledges a Research Fellowship from All Souls College, Oxford.
C.R.D. was supported in part by US National Science Foundation award DMS-1813003.
J.S.W. acknowledges support from NASA Grant NNH13ZDA001N-CRYO and Swedish Research Council grant no. 638-2013-9243. A.J.W. and C.R.D. acknowledge the hospitality of the Woods Hole Oceanographic Institution Geophysical Fluid Dynamics Program, which is supported by NSF OCE-1829864, whilst part of this work was developed.

\appendix
\section{Appendix A: Simulation details} \label{sec:convergence}
\subsection{Spatial resolution: Comparison with the Kolmogorov length scale}
Following \citet{grotzbach1983}, the Kolmogorov length scale, $\eta$, in Rayleigh-B\'enard convection can be estimated as:
\begin{equation}
\eta = \left(\frac{Pr^2}{Ra \, Nu}\right)^{1/4}.
\label{eqn:eta}
\end{equation}
To obtain equation \ref{eqn:eta}, we first take the dot product of the dimensional momentum equation with $\boldsymbol{u}$, giving:
\begin{widetext}
\begin{equation}
\frac{1}{2} \, \frac{\partial u_i^2}{\partial t} + \frac{1}{2} \, \frac{\partial (u_k \, u_i^2)}{\partial x_k} = - \frac{\partial (u_i \, p)}{\partial x_i} + \alpha \, g \, w \, T \, \delta_{i2} + \nu \left[\frac{1}{2}\frac{\partial^2 u_i^2}{\partial x_k^2} - \left(\frac{\partial u_i}{\partial x_k}\right)^2\right].
\label{eqn:eta2}
\end{equation}
\end{widetext}
Taking the long time and area average of equation \ref{eqn:eta2}, we obtain
\begin{equation}
\epsilon \equiv \nu \, \left<|\nabla \boldsymbol{u}|^2\right> = \alpha \, g \, \left<w \, T\right>.
\end{equation}
Using $\left<w \, T\right> = \epsilon/(\alpha \, g) \approx \kappa \Delta T/H \times Nu$ in the expression for the dimensional Kolmogorov length scale ($\eta = (\nu^3/\epsilon)^{1/4}$) and after some algebra and rearrangement, we obtain equation \ref{eqn:eta}.

A criterion for a simulation to be well resolved is \citep{grotzbach1983}:
\begin{equation}
N_G = \frac{\pi \, \eta}{h} > 1,
\label{eqn:K41}
\end{equation}
where $h = \sqrt{\Delta x \, \Delta z}$, where $\Delta x$ and $\Delta z$ are mesh sizes along the horizontal and vertical directions. (All length scales are non-dimensionalized using the height of the cell.) We use uniform grids in our simulations, so $\Delta x = \Delta z$ and $h = \Delta z$. In table \ref{tab:K41}, we show the computed values of $\eta$ for the six highest $Ra$ for $p = -1.5$.
\begin{widetext}
\begin{center}
\begin{table}
  \begin{center}
\def~{\hphantom{0}}
  \begin{tabular}{lccccc}
      ~~$Ra$~~  & ~~$Nu$~~   &   ~~$h$~~ & ~~ $\eta$~~ & ~~$N_{\eta} = \eta/h$~~ & ~~$N_G = \pi \eta/h$\\           [3pt]
      $2.15 \times 10^8$~~ & ~~$30.70$~~ & ~~$1.25 \times 10^{-3}$~~ & ~~$3.51 \times 10^{-3}$ ~~ & ~~$3$~~ & ~~$9$\\
      $4.64 \times 10^8$~~ & ~~40.62~~ & ~~$10^{-3}$~~ & ~~$2.70 \times 10^{-3}$ ~~ & ~~$3$~~ & ~~$9$\\
      $10^9$~~ & ~~$52.53$~~ & ~~$10^{-3}$~~ & ~~$2.10 \times 10^{-3}$~~ & ~~$2$~~ & ~~$6$\\
      $2.15 \times 10^9$~~ & ~~$70.89$~~ & ~~$8.33 \times 10^{-4}$~~ & ~~$1.60 \times 10^{-3}$~~ & ~~$2$~~ & ~~$6$ \\
      $4.64 \times 10^9$~~ & ~~$91.79$~~ & ~~$7.14 \times 10^{-4}$~~ & ~~$1.24 \times 10^{-3}$~~ & ~~$2$~~ & ~~$6$ \\
      $10^{10}$~~ & ~~$121.73$~~ & ~~$4.76 \times 10^{-4}$~~ & ~~$9.52 \times 10^{-4}$~~ & ~~$2$~~ & ~~$6$
  \end{tabular}
  \caption{Comparison of mesh size with the Kolmogorov length scale for the highest six $Ra$ and $p = -1.5$. The Kolmogorov length scale is calculated using equation \ref{eqn:eta}.}
  \label{tab:K41}
  \end{center}
\end{table}
\end{center}
\end{widetext}
It is clear from the table \ref{tab:K41} that the resolutions used by us are able to resolve the Kolmogorov length scale -- both in the interior and in the boundary layers. Note that we have used a more stringent criterion than equation \ref{eqn:K41} as we show that $N_{\eta} > 1$ as well as $N_G > 1$.

\subsection{Spatial resolution: Boundary layers}
\begin{figure}
\begin{centering}
\includegraphics[trim = 0 0 0 0, clip, width = 1\linewidth]{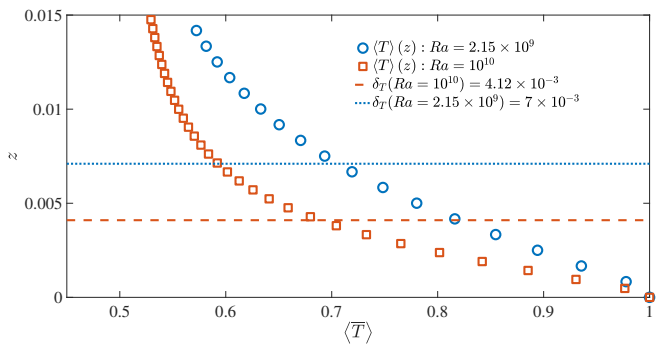} 
\caption{Horizontally and temporally averaged temperature profiles for $Ra = 2.15 \times 10^9$ (circles) and $Ra = 10^{10}$ (squares) and $p = -1.5$. The dotted and dashed lines shows the boundary-layer thicknesses for $Ra = 2.15 \times 10^9$ and $Ra = 10^{10}$, respectively. There are $9$ grid points in each boundary layer. The kinks at $z=0$ are an artifact of the use of mid-grid bounceback condition to impose no-slip and no-penetration boundary conditions in the LBM. This version of the bounceback renders the effective wall to be between the first and second grid points \citep{succi2001}. Hence, the no-slip boundary condition effectively applies at a distance $z = \Delta z/2$, where $\Delta z$ is the grid size, above $z=0$ where the temperature boundary condition is imposed. The calculation of $Nu$ at the boundary takes this into account, and has been tested by reproducing the $Nu(Ra)$ results from spectral simulations for flat boundaries \citep{TSW15b}.}
\label{fig:temp-profile}
\end{centering}
\end{figure}

We estimate the non-dimensional boundary-layer thickness, $\delta_T$, using
\begin{equation}
\delta_T = \frac{1}{2 \, Nu}.
\label{eqn:BLthick}
\end{equation}
Table \ref{tab:BL} shows the boundary-layer thickness for all $Ra$ for $p = -1.5$ and the number of grid points within the boundary layer.
\begin{widetext}
\begin{center}
\begin{table}
  \begin{center}
\def~{\hphantom{0}}
  \begin{tabular}{lcccc}
      ~~$Ra$~~  & ~~$Nu$~~   &   ~~$\Delta z$~~ & ~~ $\delta_T$~~ & ~~$N_{\delta} = \delta_T/\Delta z$~~\\           [3pt]
      $10^7$~~ & ~~$14.22$~~ & ~~$1.25 \times 10^{-3}$~~ & ~~$3.5 \times 10^{-2}$ ~~ & ~~$28$\\
      $2.15 \times 10^7$~~ & ~~15.74~~ & ~~$1.25 \times 10^{-3}$~~ & ~~$3.2 \times 10^{-2}$ ~~ & ~~$26$\\
      $4.64 \times 10^7$~~ & ~~$19.44$~~ & ~~$1.25 \times 10^{-3}$~~ & ~~$2.6 \times 10^{-2}$~~ & ~~$21$\\
      $10^8$~~ & $~~24.50$~~ & ~~$1.25 \times 10^{-3}$~~ & ~~$2.0 \times 10^{-2}$~~ & ~~$16$ \\
      $2.15 \times 10^8$~~ & ~~$30.70$~~ & ~~$1.25 \times 10^{-3}$~~ & ~~$1.6 \times 10^{-2}$ ~~ & ~~$13$\\
      $4.64 \times 10^8$~~ & ~~40.62~~ & ~~$10^{-3}$~~ & ~~$1.2 \times 10^{-2}$ ~~ & ~~$12$\\
      $10^9$~~ & ~~$52.53$~~ & ~~$10^{-3}$~~ & ~~$9.5 \times 10^{-3}$~~ & ~~$10$\\
      $2.15 \times 10^9$~~ & $~~70.89$~~ & ~~$8.33 \times 10^{-4}$~~ & ~~$7.0 \times 10^{-3}$~~ & ~~$9$ \\
      $4.64 \times 10^9$~~ & ~~91.79~~ & ~~$7.14 \times 10^{-4}$~~ & ~~$5.45 \times 10^{-3}$ ~~ & ~~$8$\\
      $10^{10}$~~ & ~~$121.73$~~ & ~~$4.76 \times 10^{-4}$~~ & ~~$4.12 \times 10^{-3}$~~ & ~~$9$\\
  \end{tabular}
  \caption{Comparison of boundary-layer thickness and the resolutions used.}
  \label{tab:BL}
  \end{center}
\end{table}
\end{center}
\end{widetext}
It is clear from the table that in our simulations there are at least $8$ grid points within the boundary layer. To further demonstrate this point, we show the time and horizontally averaged temperature profiles for $Ra = 2.15 \times 10^9$ and $Ra = 10^{10}$ and $p = -1.5$ in figure \ref{fig:temp-profile}. There are $9$ grid points in each of the two boundary layers, in agreement with the estimate in table \ref{tab:BL}.

\subsection{Temporal convergence}
\begin{figure}
\begin{centering}
\includegraphics[trim = 0 0 0 0, clip, width = 1\linewidth]{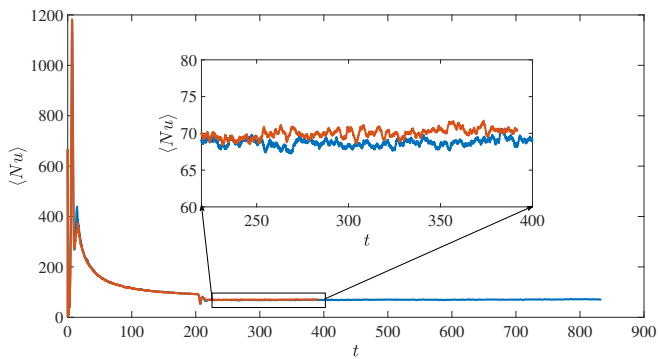} 
\caption{Moving average of the $Nu(t)$ data measured at $z/H = 0.42$ for $p = -1.5$ and $Ra = 2.15 \times 10^9$.}
\label{fig:Nu-avg1}
\end{centering}
\end{figure}

To ascertain that a time window of $200$ time units was sufficient to obtain converged statistics, we ran simulations for $Ra = 2.15 \times 10^9$ and $p = -1.5$ with the same spatial resolution for two durations: (1) $t \approx 390$ and (2) $t \approx 830$. In figure \ref{fig:Nu-avg1} we show a moving average of the $Nu(t)$ data measured at $z/H = 0.42$, where $H$ is the height of the cell for the two cases. The window for the moving average is $200$ time units.
It is seen that the moving average value of $Nu$ is approximately constant beginning at $t \approx 220$. There are fluctuations in the curves and the maximum and minimum variations in the fluctuations are about $4$\% of the mean. For the shorter duration run, which was averaged over $200$ time units, $\left<\overline{Nu}\right> = 70.89$, and for the longer duration run, which was averaged over $613$ time units, $\left<\overline{Nu}\right> = 69.81$. The difference between the two values is $1.5$ \%. 

\section{Appendix B: Characteristic length scale} \label{sec:length}
\begin{figure}  
\begin{centering} 
\includegraphics[trim = 0 0 0 0, clip, width = \linewidth]{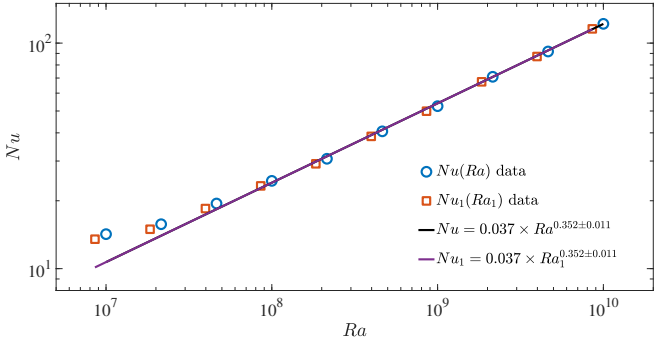} 
\caption{The figure shows $Nu_1(Ra_1)$ and $Nu(Ra)$ data sets along with their power-law fits for $p=-1.5$.}
\label{fig:NuRa_H}
\end{centering}
\end{figure}

We here consider the impact of an alternative definition of the characteristic length scale used in the non-dimensionalization. Let $H_1$ be the characteristic length scale that gives $Nu_1=1$ at $Ra = 0$. The definition of $Nu$ used above is
\begin{equation}
Nu = \frac{Q \, H}{k \, \Delta T},
\end{equation}
where $Q$ is the total heat flux, $H$ is the depth of the cell, and $\Delta T$ is the temperature difference between top and bottom boundaries. This can be written as
\begin{equation}
Nu = \frac{Q \, H_1}{k \, \Delta T} \, \frac{H}{H_1} = Nu_1 \, \frac{H}{H_1}.
\end{equation}
By design, $Nu_1 = 1$ at $Ra = 0$, and hence
\begin{equation}
\frac{H_1}{H} = \frac{1}{Nu(Ra=0)}.
\label{eqn:H}
\end{equation}
Performing simulations for $Ra = 0$ and all values of $p$ used, we find that $Nu = 1.06$ for $p = -1.5$ and $-2$ and $Nu = 1.05$ for $p=-3$. This implies that the effective length scale ($H_1$) from equation \ref{eqn:H} is $\approx 95 \%$ of the depth of the cell. We can now use this new scale, $H_1$, to calculate the re-scaled values of the Rayleigh and Nusselt numbers, which are given by:
\begin{equation}
Ra_1 = \left(\frac{H_1}{H}\right)^3 \, Ra \hspace{0.2cm} \text{and} \hspace{0.2cm} Nu_1 = \left(\frac{H_1}{H}\right) \, Nu.
\end{equation}
Figure \ref{fig:NuRa_H} shows $Nu_1(Ra_1)$, $Nu(Ra)$, and the linear least-squares fits over the last seven data points for the respective data sets for $p=-1.5$. The fit for the new data over the highest seven $Ra$ gives $Nu_1 = 0.037 \times Ra_1^{0.352 \pm 0.011}$, which is the same for the corresponding fit $Nu=0.037 \times Ra^{0.352 \pm 0.011}$ using the length scale $H$. This is easily seen in figure \ref{fig:NuRa_H}.
Similar analyses have been performed for data sets for other values of $p$, and these conclusions remain the same for those data sets as well with $H_1/H = 0.95$ for all values of $p$ considered here. Also, the pre-factor changes by less than 1\% because $H_1/H = 0.95$, which is close to unity.

Hence, although the choice of $H_1$ is relevant if one requires that the Nusselt number is $1$ in the conductive state, choosing $H$ as the length scale does not impact the scaling results reported for the turbulent heat transport for any value of $p$ used in this study in any appreciable way.

\section{Appendix C: Effect of increased area on heat transport}
\begin{figure}  
\begin{centering}
\includegraphics[scale=0.2]{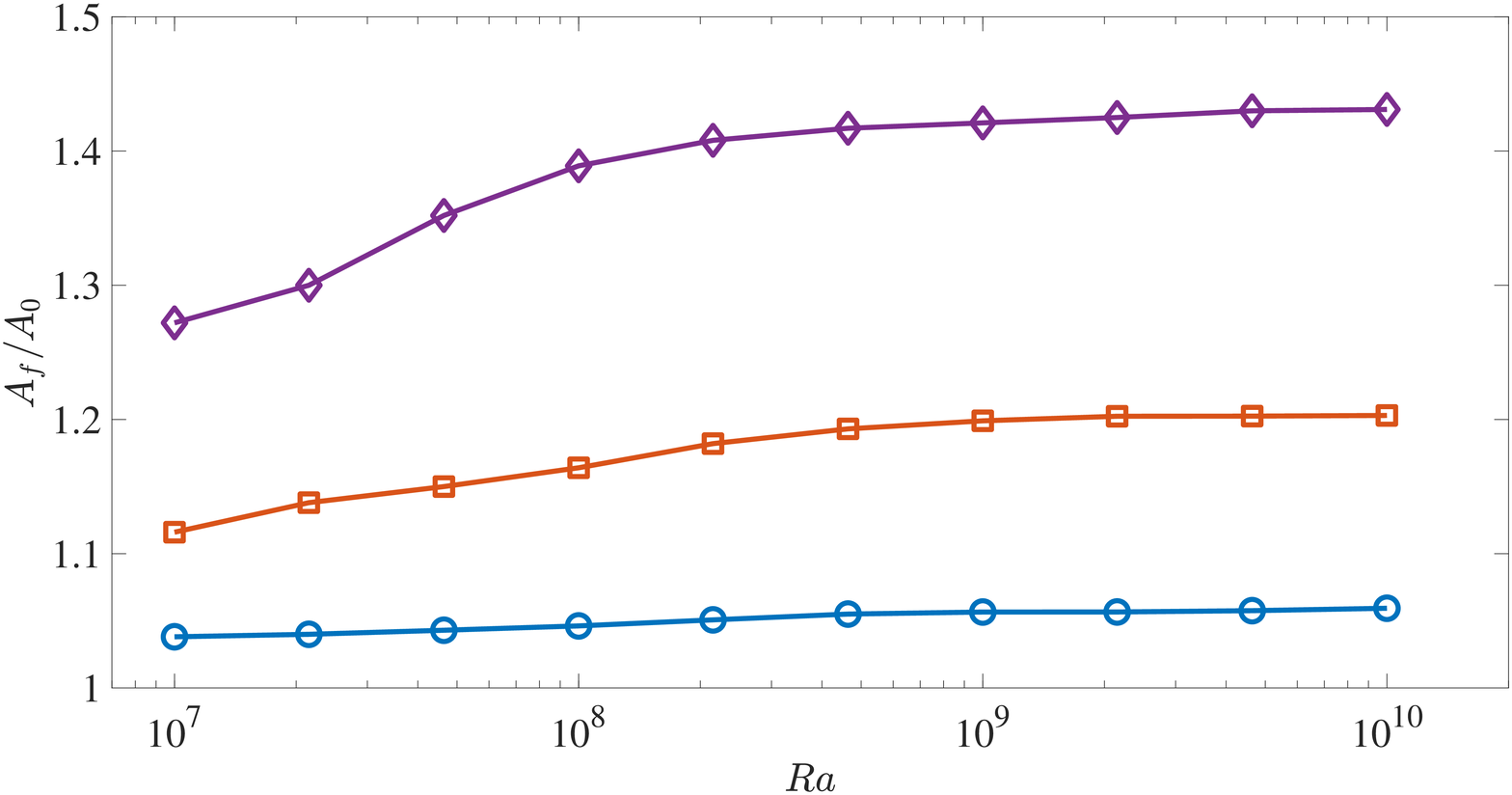} 
\caption{Ratio $A_f/A_0$ of effective transfer area to that of flat boundary as a function of $Ra$ for different values of $p$: $p=-3$ (circles); $p=-2$ (squares); and $p=-1.5$ (diamonds).}
\label{fig:EA_Ra}
\end{centering}
\end{figure}

To understand the contribution of increased surface area to the transport of heat, we compute the ratio $A_f/A_0$ (effective area) and $Nu/Nu_0$ (effective heat transport). Here, $A_f$ is the effective transfer area given by the surface area of $min\left[h(x),1-\delta_T\right]$, which increases the effective area to account for regions where the fractal boundary at $z=h(x)$ protrudes beyond the boundary layer thickness $\delta_T$; $A_0$ is the area of the flat boundary; and, $Nu_0$ is the value of the Nusselt number for a flat boundary. The boundary layer thickness is estimated using equation \ref{eqn:BLthick} and the values of $Nu_0$ are obtained using the power law: $Nu_0 = 0.138 \times Ra^{0.285}$ \citep{doering2009}. The values of $A_f/A_0$ calculated in this way are monotonic functions of $Ra$, so that as the thermal boundary layer thins, more of the fractal boundary is exposed to the flow. This is shown in figure \ref{fig:EA_Ra} for all values of $p$ (i.e., as $p$ increases from $-3$ to $-1.5$).
Note that the curves appear to saturate at the higher end of $Ra$ because we have computed the Steinhaus series \citep{rothrock1980} only up to a large finite wavenumber. For each $Ra$, the effective area increases for increasing value of $p$, showing that the boundaries become more rough.

\begin{figure}
\centering

\begin{subfigure}
\centering
\includegraphics[trim = 0 0 0 0, clip, width = 1\linewidth]{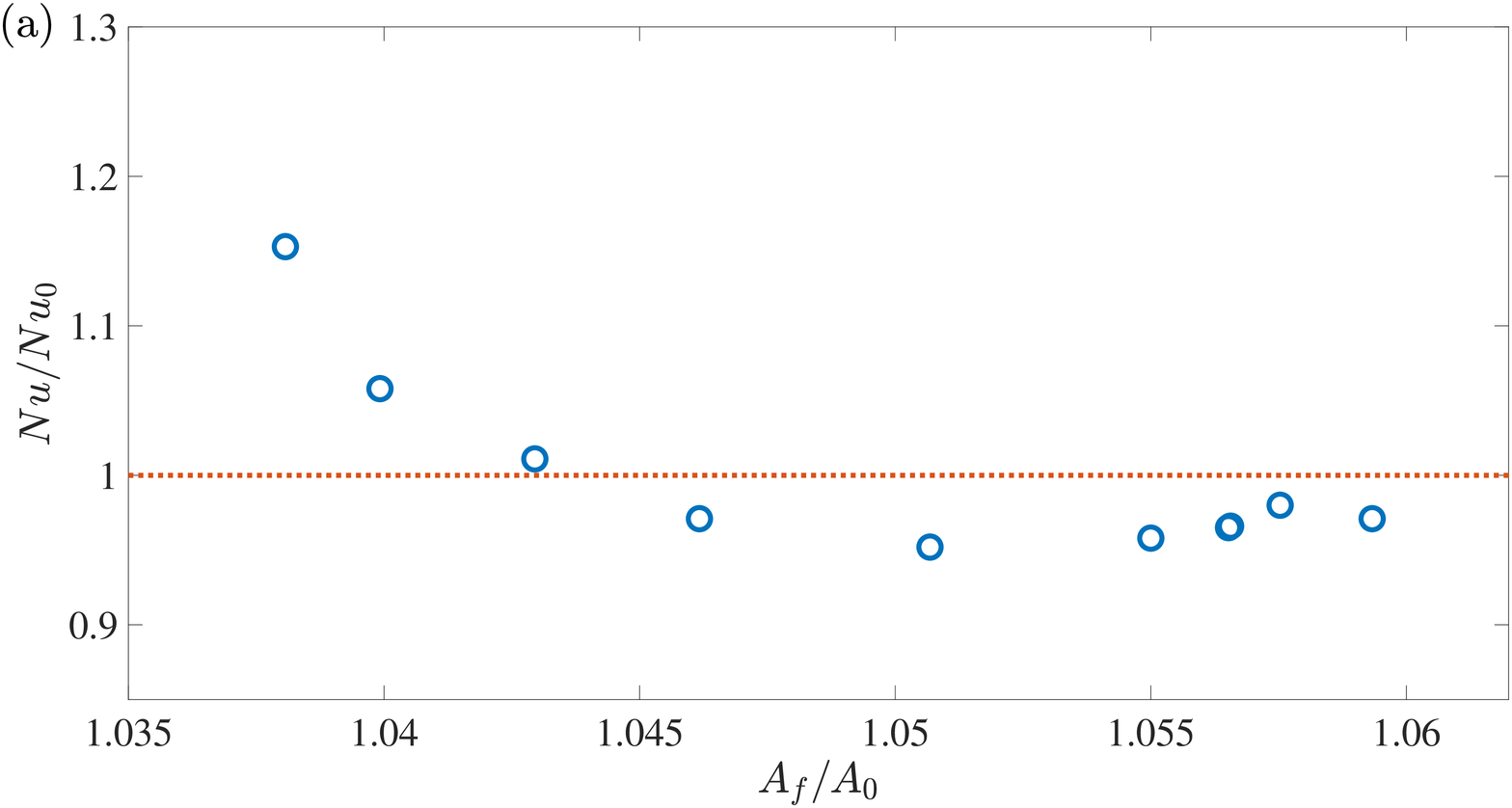}  
\end{subfigure}
    
\begin{subfigure}
\centering
\includegraphics[trim = 0 0 0 0, clip, width = 1\linewidth]{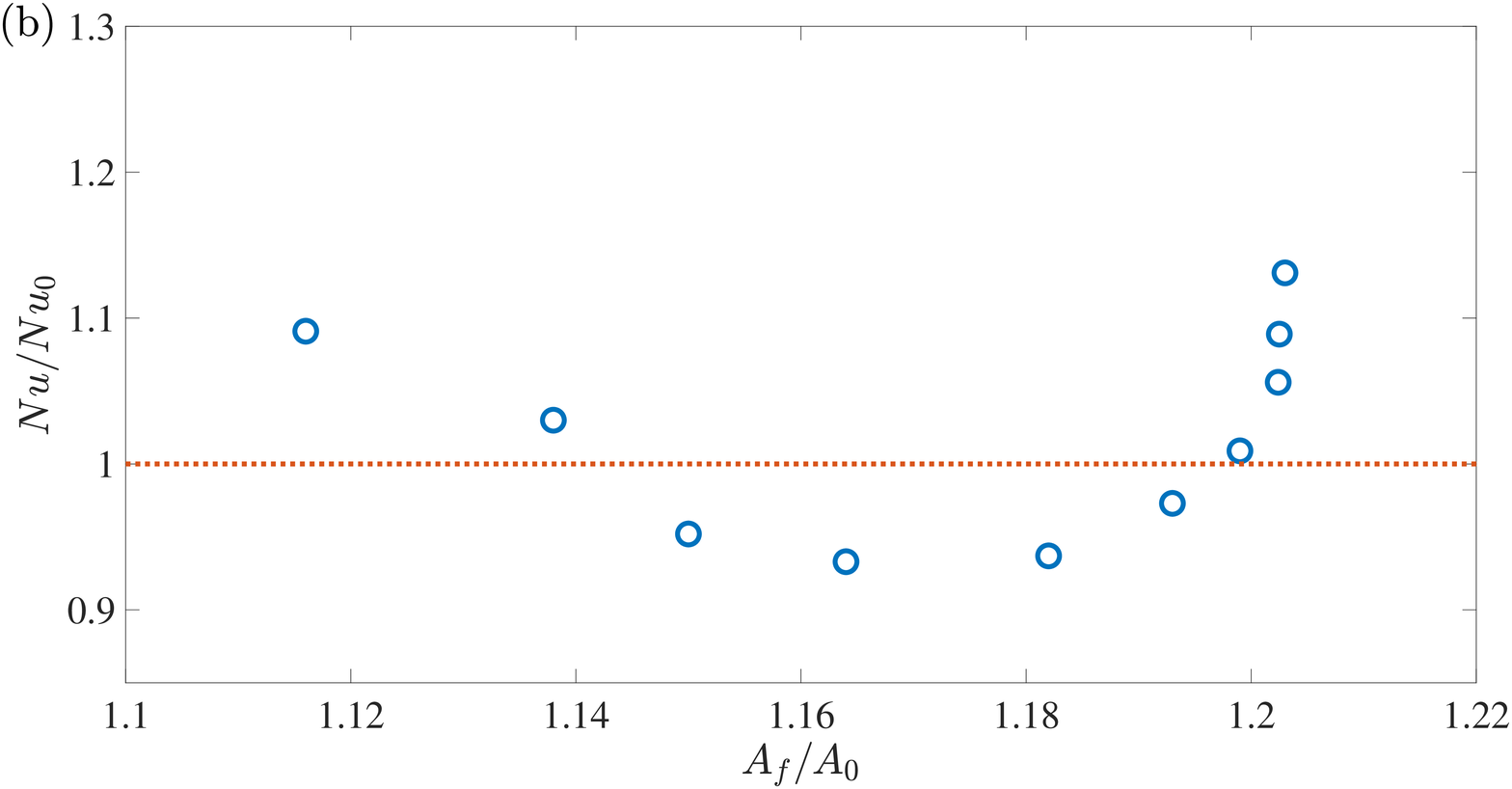} 
\end{subfigure}
          
\begin{subfigure}
\centering
\includegraphics[trim = 0 0 0 0, clip, width = 1\linewidth]{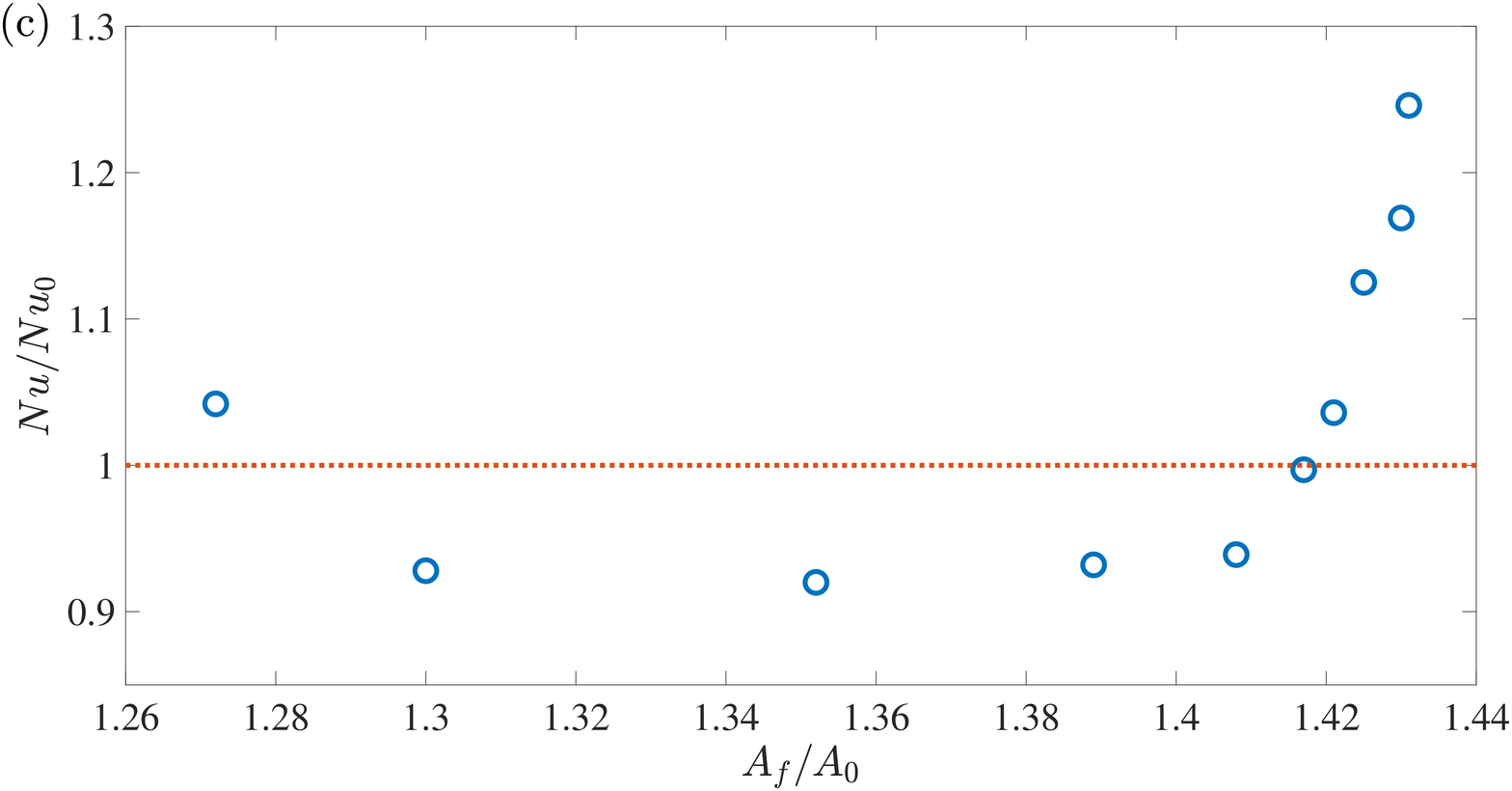}  
\end{subfigure}
          
\caption{$Nu/Nu_0$ vs. $A_f/A_0$ for: (a) $p = -3$; (b) $p=-2$; and (c) $p=-1.5$.}     
\label{fig:area}             
\end{figure}

In figures \ref{fig:area}(a)--\ref{fig:area}(c), we show $Nu/Nu_0$ as a function of $A_f/A_0$ for all $Ra$ and $p$. In each of these figures, the first point, with the lowest $A_f/A_0$, corresponds to $Ra = 10^7$, the second to $Ra = 2.15 \times 10^7$,..., and the last one, with the largest $A_f/A_0$ value, to $Ra = 10^{10}$. (See figure \ref{fig:EA_Ra}.)
The following observations can be made from figures \ref{fig:area}(a) -- \ref{fig:area}(c): (a) $Nu/Nu_0$ varies non-monotonically with $A_f/A_0$ showing the effect of the exposure of the fractal boundary to the outer flow; and (b) the difference in $A_f/A_0$ values for the last three data points for $p=-2$ and $p=-3/2$ is very small, but there is a relatively substantial increase in $Nu/Nu_0$ for these values.

\begin{figure}
\centering

\begin{subfigure}
\centering
\includegraphics[trim = 0 0 0 0, clip, width = 1\linewidth]{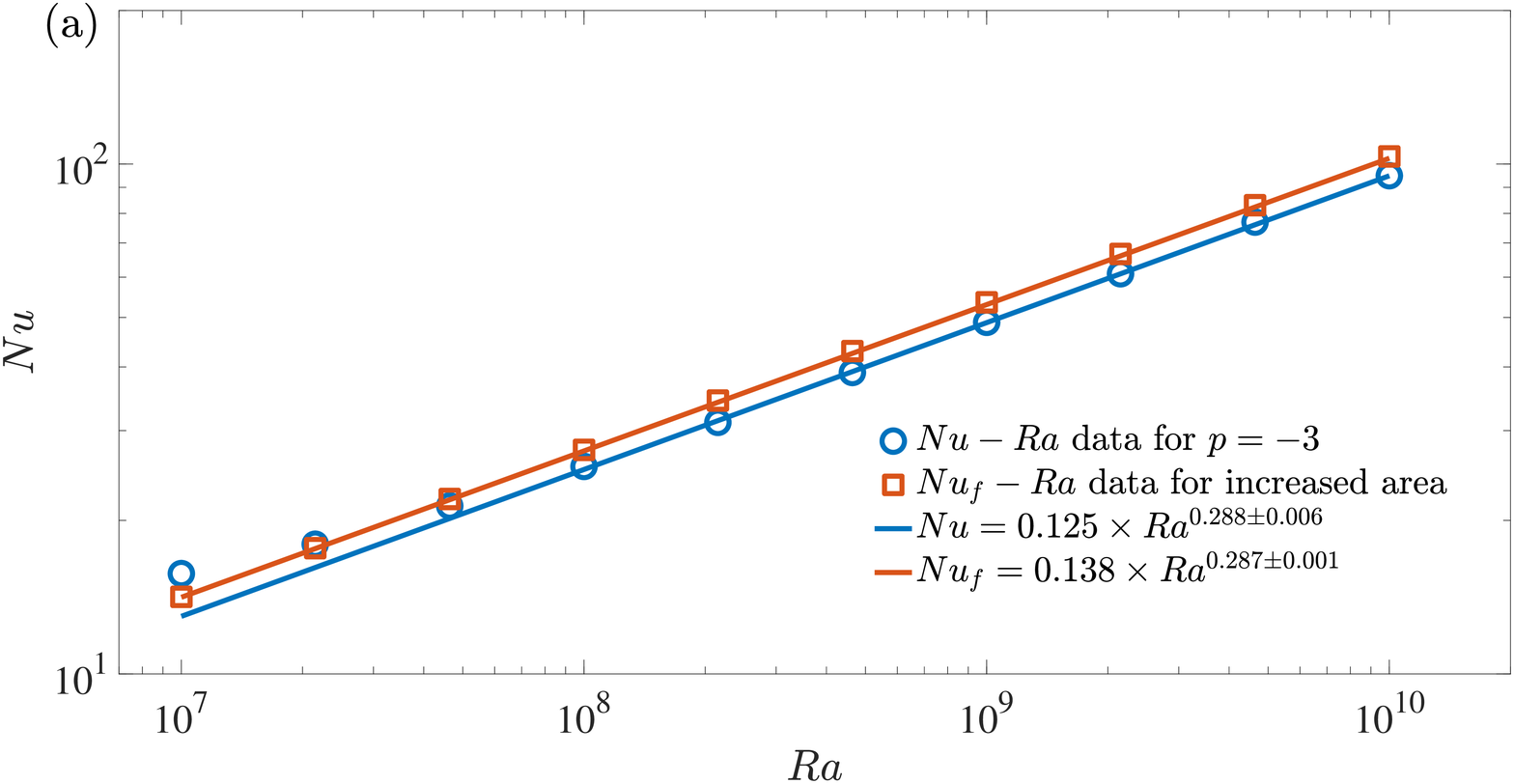}  
\end{subfigure}
    
\begin{subfigure}
\centering
\includegraphics[trim = 0 0 0 0, clip, width = 1\linewidth]{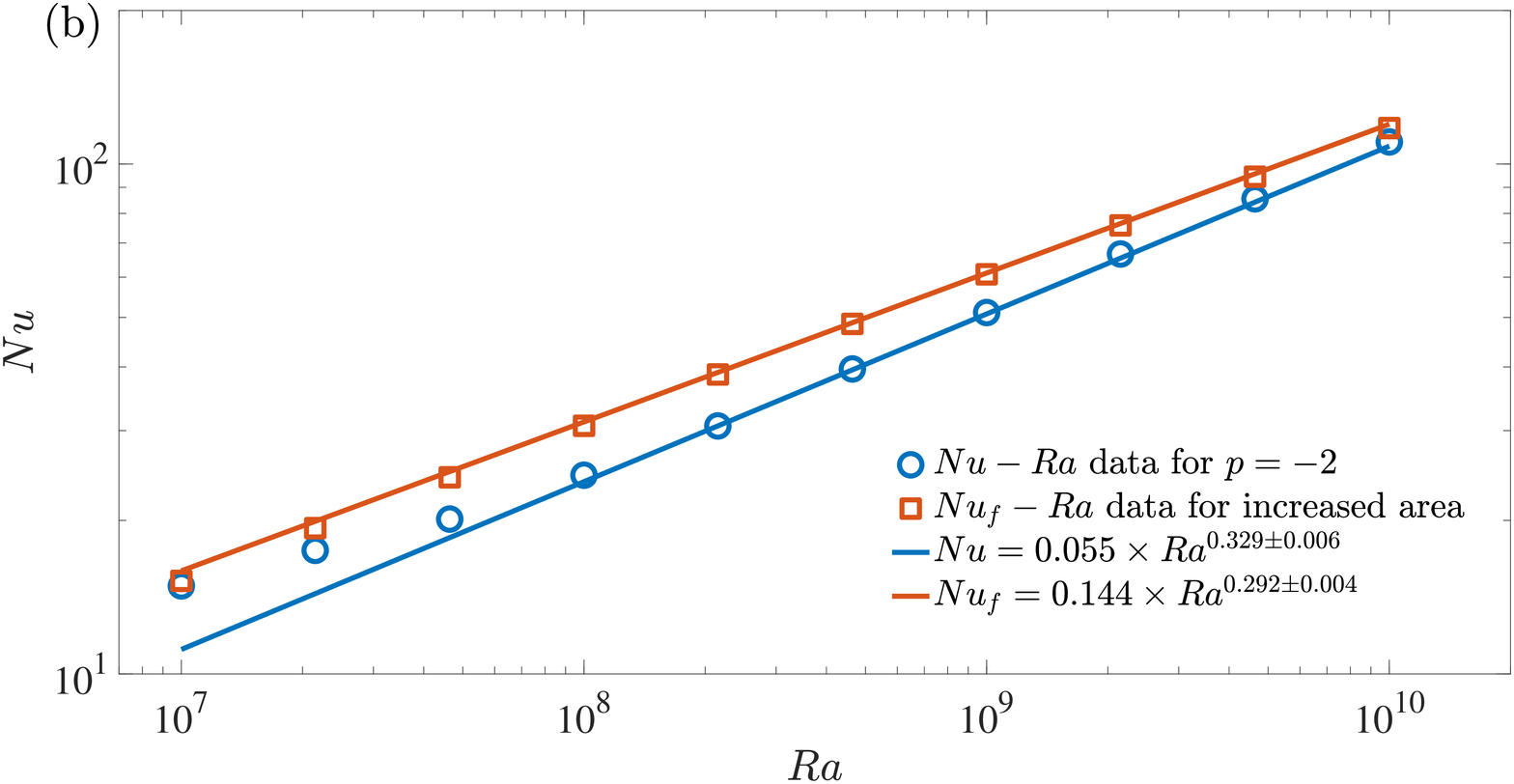} 
\end{subfigure}
          
\begin{subfigure}
\centering
\includegraphics[trim = 0 0 0 0, clip, width = 1\linewidth]{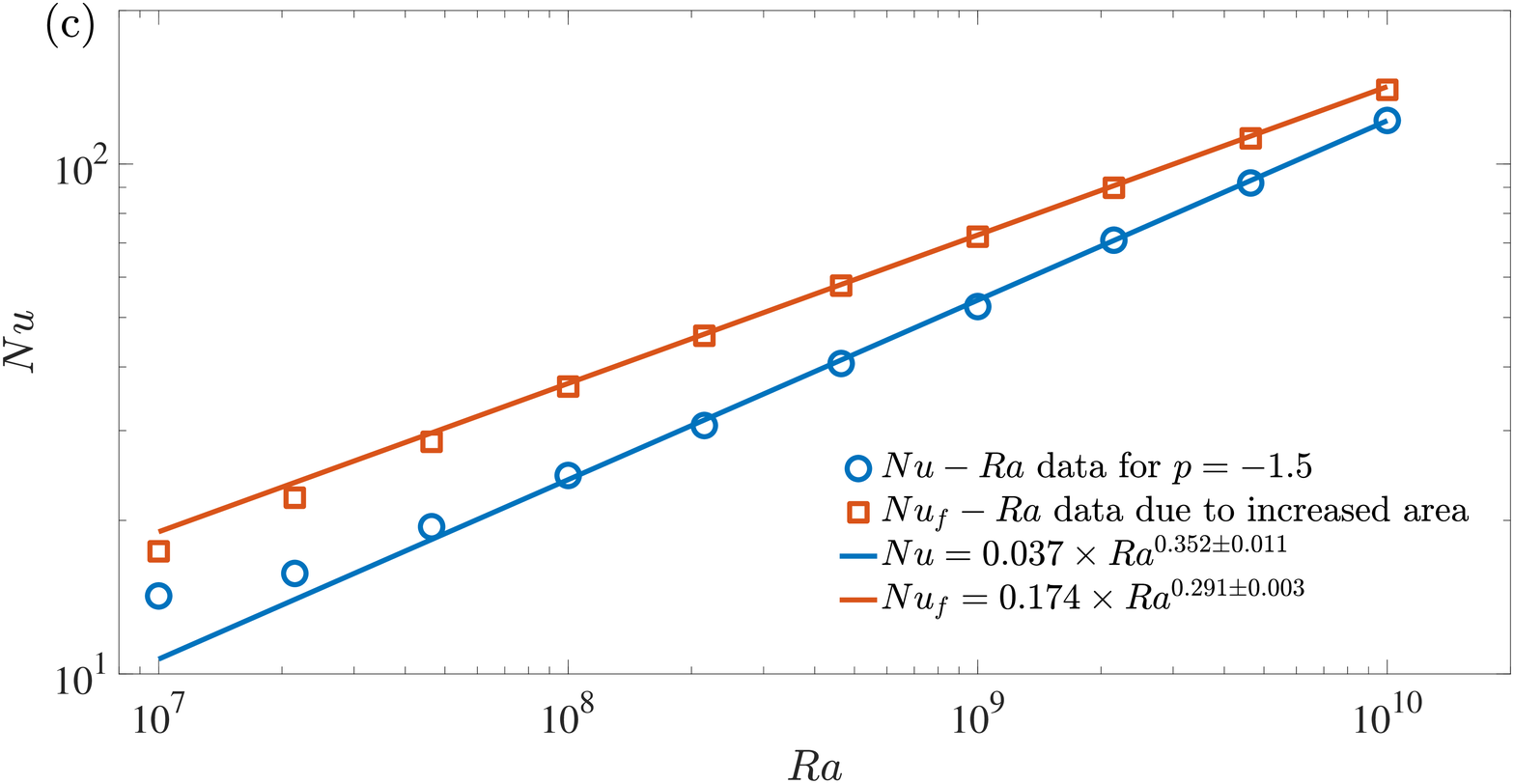}  
\end{subfigure}
          
\caption{Heat flux data for the fractal boundaries ($Nu$) and theory that applies heat flux for flat boundaries over an augmented area ($Nu_f$). (a) $p = -3$; (b) $p=-2$; and (c) $p=-1.5$. The power-law fits are for the range $Ra \in \left[10^8, 10^{10}\right]$.}     
\label{fig:area_NuRa}             
\end{figure}

To determine if the increase in the effective area can explain the augmentation in heat flux, we plot the $Nu(Ra)$ data for $p=-3, -2$, and $-1.5$ and the corresponding $Nu_f(Ra) = Nu_0(Ra) \times A_f(Ra)/A_0$ data for the same range of $Ra$ in figures \ref{fig:area_NuRa}(a)-\ref{fig:area_NuRa}(c). 
If the increase in $Nu(Ra)$ values were solely due to the enhanced area, such that
\begin{equation}
Nu = Nu_0 \, \frac{A_f}{A_0} = Nu_f,
\end{equation}
then the $Nu(Ra)$ and $Nu_f(Ra)$ curves would coincide. It is seen from figures \ref{fig:area_NuRa}(a)-\ref{fig:area_NuRa}(c) that: (a) the curves do not coincide; (b) the fractal boundary for $p=-3$ is hydrodynamically smooth for heat transport as the $Nu(Ra)$ and $Nu_f(Ra)$ curves are parallel for $Ra \ge 10^8$; (c) the curve for $Nu_f(Ra)$ does not explain the curvature that is seen at lower $Ra$ in the $Nu(Ra)$ data for any $p$; and (d) while the values of $\beta$ for the two simulated $Nu(Ra)$ curves are substantially different for $p=-2$ and $p=-1.5$, the corresponding fitted $\beta$ for $Nu_f(Ra)$ seem independent of $p$, to within the uncertainty. Hence, we conclude that the increase in $\beta$ for convection over the fractal boundaries is because of a change in the dynamics, which is consistent with increased plume production \citep{verzicco2006, TSW15b, TSW17}. This dynamics has been explicitly shown at play in previous DNS studies of turbulent convection over periodic roughness of different wavelengths \citep{TSW15b, TSW17}.

\section{Appendix D: Simulation data}
Here, we have tabulated the $Nu(Ra)$ and $Re(Ra)$ data from the simulations. The data shown in figures \ref{fig:NuRa} and \ref{fig:ReRa} correspond to Realization-4 for all values of $p$.
\begin{widetext}
\begin{center}
\begin{table}
  \begin{center}
\def~{\hphantom{0}}
  \begin{tabular}{lcccc}
      ~~$Ra$~~  & ~~$Nu, r=1$~~   &   ~~$Nu, r =2$~~ & ~~$Nu, r = 3$~~ & ~~$Nu, r = 4$~~\\[3pt]
       $10^7$~~   & ~~$14.96 \pm 0.07$~~ & ~~$14.31 \pm 0.04$~~ & ~~$15.13 \pm 0.09$~~  & ~~$15.73 \pm 0.04$~~             \\
       $2.15 \times 10^7$~~    & ~~$17.91 \pm 0.10$~~ & ~~$17.75 \pm 0.10$~~ & ~~$17.85 \pm 0.08$~~ & ~~$17.96 \pm 0.12$~~         \\
       $4.64 \times 10^7$~~  & ~~$21.10 \pm 0.23$~~ & ~~$20.75 \pm 0.15$~~ & ~~$21.01 \pm 0.06$~~   & ~~$21.35 \pm 0.03$~~\\
       $10^8$~~   & ~~$25.33 \pm 0.11$~~ & ~~$25.32 \pm 0.16$~~ & ~~$25.37 \pm 0.12$~~ & ~~$25.52 \pm 0.05$~~\\
       $2.15 \times 10^8$~~ & ~~$30.50 \pm 0.25$~~ & ~~$30.58 \pm 0.33$~~ & ~~$30.38 \pm 0.41$~~ & ~~$31.14 \pm 0.15$~~\\
       $4.64 \times 10^8$~~ & ~~$37.98 \pm 0.26$~~ & ~~$38.73 \pm 0.24$~~ & ~~$38.84 \pm 0.26$~~ & ~~$39.01 \pm 0.15$~~ \\
       $10^9$~~ & ~~$48.36 \pm 0.52$~~ & ~~$48.49 \pm 0.22$~~ & ~~$48.75 \pm 0.59$~~ & ~~$48.92 \pm 0.29$~~\\
       $2.15 \times 10^9$~~ & ~~$61.74 \pm 1.15$~~ & ~~$61.36 \pm 0.73$~~ & ~~$60.30 \pm 0.81$~~ & ~~$60.92 \pm 0.86$~~\\
       $4.64 \times 10^9$~~ & ~~$-$~~ & ~~$-$~~ & ~~$-$~~ & ~~$76.92 \pm 1.60$~~ \\
       $10^{10}$~~ & ~~$-$~~ & ~~$-$~~ & ~~$-$~~ & ~~$94.85 \pm 3.07$~~ \\
       \end{tabular}
  \caption{$Nu(Ra)$ data for four different realizations of rough boundary for $p = -3.0$. The different realizations are numbered as $r=1,..,4$.}
  \label{tab:p3_NuRa}
  \end{center}
\end{table}
\end{center}
\end{widetext}

\begin{widetext}
 \begin{center}
\begin{table}
  \begin{center}
\def~{\hphantom{0}}
  \begin{tabular}{lcccc}
      ~~$Ra$~~  & ~~$Nu, r=1$~~   &   ~~$Nu, r =2$~~ & ~~$Nu, r = 3$~~ & ~~$Nu, r = 4$~~\\[3pt]
       $10^7$~~   & ~~$13.79 \pm 0.06$~~ & ~~$13.76 \pm 0.07$~~ & ~~$14.72 \pm 0.06$~~  & ~~$14.88 \pm 0.01$~~             \\
       $2.15 \times 10^7$~~    & ~~$16.29 \pm 0.11$~~ & ~~$16.68 \pm 0.09$~~ & ~~$17.02 \pm 0.05$~~ & ~~$17.47 \pm 0.07$~~         \\
       $4.64 \times 10^7$~~  & ~~$19.77 \pm 0.05$~~ & ~~$19.92 \pm 0.15$~~ & ~~$19.73 \pm 0.12$~~   & ~~$20.12 \pm 0.06$~~\\
       $10^8$~~   & ~~$23.92 \pm 0.23$~~ & ~~$24.52 \pm 0.32$~~ & ~~$24.41 \pm 0.34$~~ & ~~$24.53 \pm 0.24$~~\\
       $2.15 \times 10^8$~~ & ~~$30.14 \pm 0.12$~~ & ~~$30.05 \pm 0.19$~~ & ~~$30.49 \pm 0.48$~~ & ~~$30.64 \pm 0.24$~~\\
       $4.64 \times 10^8$~~ & ~~$39.61 \pm 0.18$~~ & ~~$39.24 \pm 0.10$~~ & ~~$39.35 \pm 0.10$~~ & ~~$39.63 \pm 0.19$~~ \\
       $10^9$~~ & ~~$50.95 \pm 0.56$~~ & ~~$50.61 \pm 0.36$~~ & ~~$50.22 \pm 0.63$~~ & ~~$51.13 \pm 0.36$~~\\
       $2.15 \times 10^9$~~ & ~~$66.73 \pm 1.16$~~ & ~~$67.72 \pm 1.31$~~ & ~~$68.18 \pm 1.26$~~ & ~~$66.60 \pm 2.28$~~\\
       $4.64 \times 10^9$~~ & ~~$-$~~ & ~~$-$~~ & ~~$-$~~ & ~~$85.45 \pm 2.43$~~ \\
       $10^{10}$~~ & ~~$-$~~ & ~~$-$~~ & ~~$-$~~ & ~~$110.49 \pm 10.67$~~ \\
       \end{tabular}
  \caption{$Nu(Ra)$ data for four different realizations of rough boundary for $p = -2.0$. The different realizations are numbered as $r=1,..,4$.}
  \label{tab:p2_NuRa}
  \end{center}
\end{table}
\end{center}
\end{widetext}

\begin{widetext}
 \begin{center}
\begin{table}
  \begin{center}
\def~{\hphantom{0}}
  \begin{tabular}{lcccc}
      ~~$Ra$~~  & ~~$Nu, r=1$~~   &   ~~$Nu, r =2$~~ & ~~$Nu, r = 3$~~ & ~~$Nu, r = 4$~~\\[3pt]
       $10^7$~~   & ~~$14.34 \pm 0.02$~~ & ~~$13.92 \pm 0.04$~~ & ~~$14.14 \pm 0.02$~~  & ~~$14.22 \pm 0.02$~~             \\
       $2.15 \times 10^7$~~    & ~~$15.74 \pm 0.18$~~ & ~~$16.30 \pm 0.09$~~ & ~~$15.74 \pm 0.17$~~ & ~~$15.74 \pm 0.04$~~         \\
       $4.64 \times 10^7$~~  & ~~$19.31 \pm 0.08$~~ & ~~$19.95 \pm 0.15$~~ & ~~$19.42 \pm 0.10$~~   & ~~$19.44 \pm 0.08$~~\\
       $10^8$~~   & ~~$24.34 \pm 0.34$~~ & ~~$24.11 \pm 0.34$~~ & ~~$24.22 \pm 0.25$~~ & ~~$24.50 \pm 0.14$~~\\
       $2.15 \times 10^8$~~ & ~~$29.95 \pm 0.31$~~ & ~~$31.02 \pm 0.32$~~ & ~~$30.85 \pm 0.33$~~ & ~~$30.70 \pm 0.42$~~\\
       $4.64 \times 10^8$~~ & ~~$40.17 \pm 0.17$~~ & ~~$40.79 \pm 0.16$~~ & ~~$40.94 \pm 0.09$~~ & ~~$40.62 \pm 0.30$~~ \\
       $10^9$~~ & ~~$53.05 \pm 0.41$~~ & ~~$53.15 \pm 0.32$~~ & ~~$54.78 \pm 0.26$~~ & ~~$52.53 \pm 0.37$~~\\
       $2.15 \times 10^9$~~ & ~~$69.77 \pm 0.79$~~ & ~~$71.96 \pm 1.60$~~ & ~~$71.28 \pm 1.02$~~ & ~~$70.89 \pm 1.52$~~\\
       $4.64 \times 10^9$~~ & ~~$-$~~ & ~~$-$~~ & ~~$-$~~ & ~~$91.79 \pm 1.41$~~ \\
       $10^{10}$~~ & ~~$-$~~ & ~~$-$~~ & ~~$-$~~ & ~~$121.73 \pm 11.52$~~ \\
       \end{tabular}
  \caption{$Nu(Ra)$ data for four different realizations of rough boundary for $p = -1.5$. The different realizations are numbered as $r=1,..,4$.}
  \label{tab:p32_NuRa}
  \end{center}
\end{table}
\end{center}
\end{widetext}

\begin{widetext}
 \begin{center}
\begin{table}
  \begin{center}
\def~{\hphantom{0}}
  \begin{tabular}{lcccc}
      ~~$Ra$~~  & ~~$Re, r=1$~~   &   ~~$Re, r =2$~~ & ~~$Re, r = 3$~~ & ~~$Re, r = 4$~~\\[3pt]
       ~~$10^7$~~   & ~~$941$~~ & ~~$930$~~ & ~~$924$~~ & ~~$938$~~\\
       ~~$2.15 \times 10^7$~~    & ~~$1421$~~ & ~~$1382$~~ & ~~$1397$~~ & ~~$1449$~~ \\
       ~~$4.64 \times 10^7$~~  & ~~$2133$~~ & ~~$2129$~~ & ~~$2147$~~ & ~~$2197$~~ \\
       ~~$10^8$~~   & ~~$3307$~~ & ~~$3282$~~ & ~~$3296$~~ & ~~$3283$~~\\
       ~~$2.15 \times 10^8$~~ & ~~$5336$~~ & ~~$5692$~~ & ~~$5551$~~ & ~~$5213$~~\\
       ~~$4.64 \times 10^8$~~ & ~~$8743$~~ & ~~$9154$~~ & ~~$8710$~~ & ~~$9107$~~\\
       ~~$10^9$~~ & ~~$13081$~~ & ~~$13582$~~ & ~~$13742$~~ & ~~$13661$~~\\
       ~~$2.15 \times 10^9$~~ & ~~$21317$~~ & ~~$19841$~~ & ~~$19465$~~ & ~~$21153$~~\\
       ~~$4.64 \times 10^9$~~ & ~~$-$~~ & ~~$-$~~ & ~~$-$~~ & ~~$30781$~~ \\
       ~~$10^{10}$~~ & ~~$-$~~ & ~~$-$~~ & ~~$-$~~ & ~~$44723$~~\\
  \end{tabular}
  \caption{$Re(Ra)$ data for four different realizations of rough boundary for $p = -3.0$. The different realizations are numbered as $r=1,..,4$.}
  \label{tab:kd}
  \end{center}
\end{table}
\end{center}
\end{widetext}

\begin{widetext}
 \begin{center}
\begin{table}
  \begin{center}
\def~{\hphantom{0}}
  \begin{tabular}{lcccc}
      ~~$Ra$~~  & ~~$Re, r=1$~~   &   ~~$Re, r =2$~~ & ~~$Re, r = 3$~~ & ~~$Re, r = 4$~~\\[3pt]
       ~~$10^7$~~   & ~~$927$~~ & ~~$937$~~ & ~~$889$~~  & ~~$909$~~\\
       ~~$2.15 \times 10^7$~~    & ~~$1461$~~ & ~~$1421$~~ & ~~$1371$~~ & ~~$1385$~~ \\
       ~~$4.64 \times 10^7$~~  & ~~$2232$~~ & ~~$2154$~~ & ~~$2292$~~ & ~~$2162$~~ \\
       ~~$10^8$~~   & ~~$3482$~~ & ~~$3503$~~ & ~~$3449$~~ & ~~$3551$~~\\
       ~~$2.15 \times 10^8$~~ & ~~$5587$~~ & ~~$5618$~~ & ~~$5414$~~ & ~~$5633$~~\\
       ~~$4.64 \times 10^8$~~ & ~~$9060$~~ & ~~$9001$~~ & ~~$8772$~~ & ~~$9317$~~\\
       ~~$10^9$~~ & ~~$13478$~~ & ~~$13149$~~ & ~~$13492$~~ & ~~$14369$~~\\
       ~~$2.15 \times 10^9$~~ & ~~$20794$~~ & ~~$21062$~~ & ~~$21169$~~ & ~~$20735$~~\\
       ~~$4.64 \times 10^9$~~ & ~~$-$~~ & ~~$-$~~ & ~~$-$~~ & ~~$31900$~~  \\
       ~~$10^{10}$~~ & ~~$-$~~ & ~~$-$~~ & ~~$-$~~ & ~~$44341$~~\\
  \end{tabular}
  \caption{$Re(Ra)$ data for four different realizations of rough boundary for $p = -2.0$. The different realizations are numbered as $r=1,..,4$.}
  \label{tab:kd}
  \end{center}
\end{table}
\end{center}
\end{widetext}

\begin{widetext}
 \begin{center}
\begin{table}
  \begin{center}
\def~{\hphantom{0}}
  \begin{tabular}{lcccc}
      ~~$Ra$~~  & ~~$Re, r=1$~~   &   ~~$Re, r =2$~~ & ~~$Re, r = 3$~~ & ~~$Re, r = 4$~~\\[3pt]
       ~~$10^7$~~   & ~~$893$~~ & ~~$881$~~ & ~~$874$~~  & ~~$892$~~\\
       ~~$2.15 \times 10^7$~~    & ~~$1435$~~ & ~~$1340$~~ & ~~$1414$~~ & ~~$1451$~~ \\
       ~~$4.64 \times 10^7$~~  & ~~$2190$~~ & ~~$2201$~~ & ~~$2168$~~ & ~~$2253$~~ \\
       ~~$10^8$~~   & ~~$3536$~~ & ~~$3573$~~ & ~~$3552$~~ & ~~$3458$~~\\
       ~~$2.15 \times 10^8$~~ & ~~$5695$~~ & ~~$5786$~~ & ~~$5575$~~ & ~~$5772$~~\\
       ~~$4.64 \times 10^8$~~ & ~~$9124$~~ & ~~$8819$~~ & ~~$8934$~~ & ~~$8845$~~\\
       ~~$10^9$~~ & ~~$13302$~~ & ~~$13273$~~ & ~~$13097$~~ & ~~$13857$~~\\
       ~~$2.15 \times 10^9$~~ & ~~$20178$~~ & ~~$20849$~~ & ~~$20765$~~ & ~~$20628$~~\\
       ~~$4.64 \times 10^9$~~ & ~~$-$~~ & ~~$-$~~ & ~~$-$~~ & ~~$32237$~~  \\
       ~~$10^{10}$~~ & ~~$-$~~ & ~~$-$~~ & ~~$-$~~ & ~~$44575$~~\\
  \end{tabular}
  \caption{$Re(Ra)$ data for four different realizations of rough boundary for $p = -1.5$. The different realizations are numbered as $r=1,..,4$.}
  \label{tab:kd}
  \end{center}
\end{table}
\end{center}
\end{widetext}

\bibliography{fractal.bib}

\end{document}